\documentclass{aa}
\usepackage{graphicx}
\usepackage{txfonts}
\usepackage{lipsum}
\usepackage{subcaption}         % necessary for continued figures, example in section 3
\usepackage{lscape}             % to rotate a single page table, example in appendix.
\usepackage{placeins}           % useful with \FloatBarrier, to keep
\begin{document}

\title{Quasars behind the disk of M\,31 galaxy}
%   \subtitle{Quasars behind M\,31}

\author{P. Nedialkov\inst{1}, B.F. Williams\inst{2}, V.D. Ivanov\inst{3},
A. Valcheva\inst{1}, Y. Solovyeva\inst{4}, A. Vinokurov\inst{4},
E. Malygin\inst{4}, D. Oparin\inst{4} \and O. Sholukhova\inst{4}
%\fnmsep\thanks{Corresponding author.}
}

\institute{Department of Astronomy, Faculty of Physics, Sofia University
``St. Kliment Ohridski'', 5 J. Bourchier blvd, Sofia 1164, Bulgaria\\
\email{japet@phys.uni-sofia.bg}
%             \thanks{Shows the usage of elements in the author field}
\and Department of Astronomy, Box 351580, University of Washington,
Seattle, WA 98195, USA
\and European Southern Observatory, Karl-Schwarszchild-Str. 2, D-85748
Garching bei M\"unchen, Germany
\and Special Astrophysical Observatory, Nizhny Arkhyz 369167, Russia\\
}

\date{Accepted Feb 16, 2026}

\abstract
% context heading (optional)
% {} leave it empty if necessary
%{Quasars behind nearby galaxies help to probe the chemical
%content of their gaseous component and facilitate the evaluation
%of their proper motion.}
{}
% aims heading (mandatory)
{We aim to increase the limited number of quasars behind M\,31,
necessary for probing the chemical content of the gas and for proper
motion reference, with reliable and homogeneous redshift measurements
from emission lines.}
% methods heading (mandatory)
{ We carried out spectroscopic follow up of 32 quasar candidates.}
% results heading (mandatory)
{ We confirm 23 quasars: two are new discoveries (J004029.727+403705.68
and J004215.489+412031.52) and the rest were reported elsewhere, but
with somewhat deficient analysis; 16 spectra are published for the
first time. We report new homogeneous redshifts for 34 quasars (from
40 spectra, adding 17 from archives) and summarize all available
information about bona-fide quasars with reliable redshift, bringing
their number to 124 within the $\mu_B$=26$^m$/$\Box\arcsec$ isophote.
We carried out a comparison of redshifts from different sources and
excluded some objects with redshifts derived from low-resolution
spectra. We derive the reddening for them from the color excess with
respect to dereddened counterparts with similar redshifts in the
field. Comparisons of our reddenings with M\,31 reddening maps found
no significant correlations.}
{Most QSOs behind M\,31 show low reddening and do not probe
high-extinctions underlining the need to identify fainter quasars
behind nearby galaxies, especially behind higher extinction
regions -- probably due to a bias towards following up brighter
and  less extinct candidates. Finally, the redshifts derived
from low-resolution spectra must be treated with caution, because
they can contain significant errors.}

\titlerunning
{Quasars behind M\,31}
\authorrunning{Nedialkov et al.}
\maketitle

%%%%%%%%%%%%%%%%%%%%%%%%%%%%%%%%%%%%%%%%%%%%%%%%%%%%%%%%%%%%%%
\section{Introduction} %\lipsum[1]

Quasars behind nearby galaxies are of great value for probing
their kinematics and interstellar medium. They establish a
non-moving reference system for astrometric studies
\citep{2006ApJ...652.1213K,2013ApJ...764..161K,2019ApJ...872...24V}.
The quasars are background sources that allow to study the
extinction in the intervening absorbers that can be both nearby
low-redshift galaxies or distant high-redshift objects
\citep{1997AJ....114.2353C,2006MNRAS.367..945Y,2010A&A...512A...1M}.
The imprinted narrow absorption features from the interstellar
material (ISM) on the quasar spectra can be used to measure the
chemical composition and the velocity field of the ISM
\citep[e.g.,][]{2013MNRAS.428.2198S,2013ApJ...772..110F,2013ApJ...772..111R,2014ApJ...787..147F,2021A&A...648A.116C}.
These observations, carried out in the UV, where the strong
absorption features of the cold interstellar gas are, allow to
trace the chemical enrichment history of the intervening galaxies,
and to trace the origin of the extended gaseous structures, such
as the Magellanic stream, to a particular galaxy, in this case --
to the Small Magellanic cloud. UV bright quasars are necessary for
such studies and they are usually found from surveys in the region
such as GALEX \citep[e.g.,][]{2011ApJ...728...23W}, but they are
relatively rare and only a few exist behind the nearest galaxies.

The mere identification of quasars is a challenge on its own. Usually,
it is based on their optical colors \citep{2015ApJ...811...95P},
variability properties \citep{1999MNRAS.306..637G,2020MNRAS.499.2245C},
or their X-ray, UV or mid-infrared (mid-IR) emission \citep[][among
others]{2013ApJ...775..119C,2021ApJS..253....4M,2018A&A...618A.144G}.
The presence of a galaxy in front of the quasar introduces additional
complications -- contamination from foreground sources that in the
general case can be variable, e.g. seeing dependent or due to the
intrinsic variability of the contaminating source. However, these
problems are solvable, for example, with a combination of some of
these methods: \citet{2013A&A...549A..29C} and
\citet{2016A&A...588A..93I,2024A&A...687A..16I} easily reached a
success rate of $>$70\% despite the contamination from the
Magellanic clouds, selecting quasar candidates from near-infrared
(near-IR) colors and variability. They defined a locus of previously
known quasars on the $J$$-$$K_S$ vs $Y$$-$$J$ color-color diagram
with photometry from the VISTA Magellanic Survey
\citep[VMC;][]{2011A&A...527A.116C} and followed up spectroscopically
objects located inside this region. Its borders were adjusted to
reduce the contamination, mostly from evolved stars. This is
acceptable as long as the goal of the search is to efficiently
identify quasars, but not necessarily to obtain a complete sample.
Next, they took advantage of the time-series data from the VMC survey
to identify the most variable objects by parameterizing the light
curves with a simple linear fit and required a minimum absolute
slope value. M\,31 and M\,33 present a greater challenge than the
Magellanic system due to the higher surface stellar density, the
worse contamination by objects in the more crowded intervening
galaxies and because of the generally higher extinction, except
for the inner SMC region.

Here we report the confirmation of 23 quasar candidates behind
the M\,31 disk with spectra (resolution
{700$\le$$\lambda$/$\Delta\lambda$$\le$1100}).
%; \textbf{cyan, yellow or} black solid dots in Fig.\,\ref{fig:qso_before_GDR3.png
Two of them are completely
new discoveries, and the other 21 spectra confirm the quasar nature
of previously announced candidates. We also uniformly measured
redshifts for 34 unique quasars from 40 spectra -- adding 17
archival to our data. Summarizing, we built a list of 125
spectroscopically confirmed QSOs within the
$\mu_B$=26$^m$/$\Box\arcsec$ isophote (Sec.\,\ref{sec:selection}).

The next two sections describe out target selection and the observations.
Section\,\ref{sec:quasar_sample} presents the analysis of the spectra,
Section\,\ref{sec:extinction} is an extinction study and \ref{sec:summary}
gives a summary.

\section{Target selection}\label{sec:selection}

M\,31 and M\,33 present a greater challenge than the Magellanic
clouds due to the higher surface stellar density. The census of
the quasars behind any galactic disk depends on its adopted
boundaries. For example, \citet{2019AJ....157..227M}
list 8 previously known and 7 newly found QSOs\footnote{Here,
we use quasar, QSO and AGN interchangeably, because determining
the exact energetics of each object is beyond the scope of this
work. Furthermore, the observed broad emission lines indicate
that our sample is dominated by Type 1 QSOs.} within the
footprint of M\,31 Local Group Galaxy Survey
\citep[LGSS;][]{2016AJ....152...62M}.
Three more QSO, known at the time, were missing in this list:
one reported by \citet{2013AJ....145..159H} and two others -- by
\citet{2019RAA....19...29L},
bringing the total number of the known QSOs within this survey
up to 18.

However, the LGSS is designed to cover the region of active star
formation, especially the eastern side that is further from the
Milky Way; the M\,31's disk extends further outside the
survey's footprint. To compensate for this offset, we adopt here
a wider and physically defined area of interest adopting as a
galaxy limit the ellipse with a major diameter D26=225$\arcmin$,
corresponding to the surface brightness level
$\mu_B$=26$^m$/$\Box\arcsec$ of \citet{1987A&AS...69..311W}. We
adopt a position angle PA=35 deg and major-to-minor diameters
ratio of 3.09 that RC3 \citep[Third Reference Catalog of Bright
Galaxies;][]{1991rc3..book.....D} reports for the somewhat
brighter, but better defined standard isophote of
$\mu_B$=25$^m$/$\Box\arcsec$.
For the center of M\,31, we adopt the coordinates as determined
in 2MASS \citep{2006AJ....131.1163S}:
$\alpha$=0:42:44.33 $\delta$=+41:16:07.5 (J2000).

Our ellipse encompasses
well all footprints of the modern catalogs with IR and optical
photometry in M\,31, as well as the dust maps available in the
literature.

We found spectra of 6 candidates, in addition to the previously
known 18 quasars within the selected area. Four of these extra spectra
were observed by LAMOST (Large Sky Area Multi-Object Fibre
Spectroscopic Telescope) as a part of its Quasar Survey 2012-2017
\citep{2018AJ....155..189D, 2019RAA....19...29L} and are publicly
available in LAMOST
DR5\footnote{\url{http://dr5.lamost.org/v3/sas/fits/}}. Other 2
QSOs are reported by \citet{2013AJ....145..159H}. Interestingly, the
latest LAMOST Quasar Survey DR6-DR9 2017-2021
\citep{2023ApJS..265...25J} does not report any QSOs behind the
M\,31 disk.

\begin{figure}[t!]
\centering
\includegraphics[width=1.0\hsize]{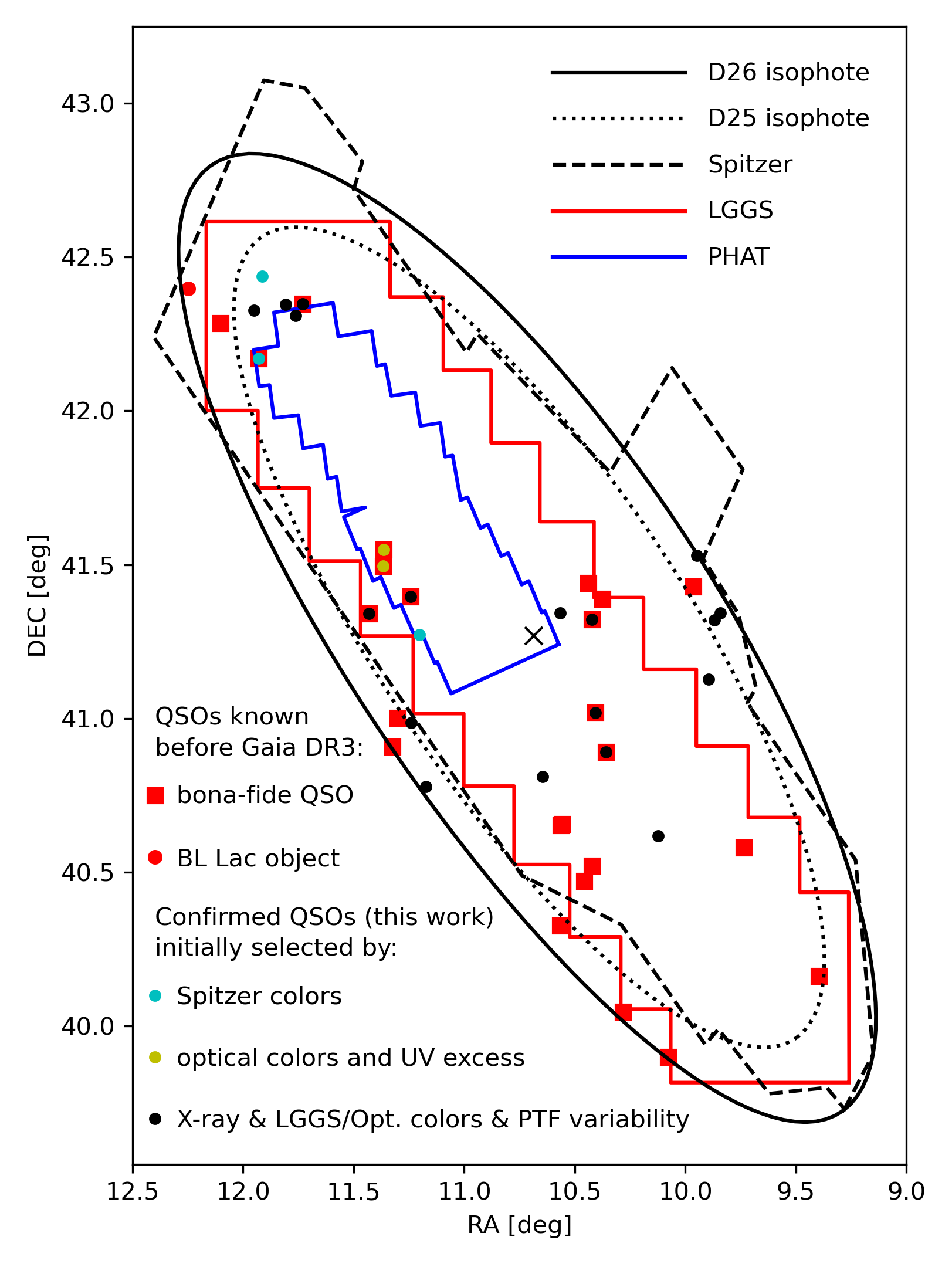}
\caption{Map of the 24 bona-fide QSOs (red squares) with
prominent emission lines and 1 BL\,Lac object (red dot) known
from the literature before Gaia\,DR3 \citep{2023A&A...674A...41A}.
The confirmed QSOs from our target list are also shown: 3 {\it
Spitzer}-selected (cyan), 2 selected from optical colors and UV
excess (yellow) and 18 selected by other means (see
Table\,\ref{tab:obslog}; black). All objects are located within
the $\mu_B$=26$^m$/$\Box\arcsec$ isophote (black solid line).
The $\mu_B$=25$^m$/$\Box\arcsec$ isophote (dotted line), the
approximate boundaries of the {\it Spitzer} PSC catalog (dashed
line) and the footprints of LGGS (red) and PHAT (blue) catalogs
are also plotted.}
\label{fig:qso_before_GDR3.png}
\end{figure}

In total, there are 25 objects with spectroscopically derived
optical redshifts within the D26 ellipse: 24 bona-fide QSOs with
prominent emission lines and a BL\,Lac discovered by
\citet{2017ApJ...851..135P}. Their positions are shown in
Fig.\,\ref{fig:qso_before_GDR3.png}, together with D26 and D25
isophotes and the footprints of M\,31 photometric catalogs:
\citet{2016AJ....152...62M}, \citet[PHAT;][]{2014ApJS..215....9W}
and \citet[{\it Spitzer} PSC;][] {2017ApJS..228....5K}.

Recently, Gaia DR3 contributed to the global census of QSOs
adding 6.4 million sources classified as AGN \citep{2023A&A...674A...31A}.
The extragalactic catalog of Gaia DR3 \citep{2023A&A...674A...41A}
reports parameters derived from various post-processing modules
dedicated to the classification and characterization of more than
6.64 million objects considered as QSO candidates but only 159
quasar candidates behind the M\,31 disk are classified as AGN.
Among them 120 have derived redshifts based on the spacecraft's
spectrograph with a resolving power R$\sim$30-100. However, a
careful comparison with better quality data from the literature,
including our own measurements reported here, suggested that
about 30\% of these redshifts are significantly different from
measurements that we deem more reliable.
This broadly agrees with their own assessment that $\sim$36\%
of the quasars have redshift errors $>$0.1\,dex, and of
Quaia survey\footnote{\url{https://zenodo.org/records/10403370}}
that finds an even higher fraction of $\sim$47\% (if the warning
flags are ignored). Therefore, we omitted
from our analysis these Gaia redshifts. Objects included in this
list are still considered, if other sources reported redshifts,
and those other redshifts were used.

\citet{2023ApJ...944....1D} listed 184 ``quasars behind the
disk of M\,31'' from spectroscopy with Dark Energy Spectroscopic
Instrument \citep[DESI;][]{2024AJ....168...58D} but only 14 of
them are located within D26 ellipse. They are by-product of
comprehensive spectral survey (resolution
2000$\le$$\lambda$/$\Delta\lambda$$\le$5500{\bf )} of 11,416 objects
on the Mayall 4m telescope at KPNO in two fields with
$3.\!\!{^\circ}2$ diameter, one centered on M\,31 itself and the
other to the South of it. The selection criteria involve negligible
parallaxes and proper motions and combine Gaia DR2 and {\it unWise}
photometry \citep{2014AJ....147..108L,2019ApJS..240...30S}.
They do not list errors but 51 of their quasars--many located
far outside the disk of M\,31--have redshifts in
\citet{2001AJ....121.2843B}, \citet{2012ApJ...759...11N},
\citet{2013AJ....145..159H}, \citet{2018AJ....155..189D},
\citet{2019AJ....157..227M}, and \citet{2019RAA....19...29L}. Their
object No. 141 with $z$=1.778 is classified based on the LAMOST
spectrum as a star at the velocity of M\,31, and for their object
No. 39 the LAMOST redshift is 1.600, as opposed to their 2.956;
for the remaining objects the mean redshift difference is
0.004$\pm$0.016 and we consider this a reliable source of confirmed
quasars.

\citet[][]{2024ApJ...964...69S} revisited the Gaia spectra,
also combining them with {\it unWISE} colors. They implemented cuts
based on proper motions and colors, thereby reducing the number of
contaminants by approximately fourfold. They used machine learning
to improve the Gaia redshift estimates. Although based on the same
low-resolution Gaia spectra, the number of catastrophic outliers
were reduced by a factor of 3 with respect to the unvetted analysis;
91\% of their redshifts agree with the SDSS DR16 measurements
\citep{2020ApJS..250....8L} to within 0.2\,dex. The final catalog
contains 1,295,502 quasars with G$<$20.5\,mag, and 755,850 candidate
objects in an even cleaner G$<$20.0\,mag sample. They reported 105
QSOs within the M\,31 disk. Their counterparts in Gaia\,DR3 which
have a Class AGN are 54 and the remaining 50 are not classified.

Our data and the literature reports produce a sample of 124
unique bona-fide spectroscopically confirmed quasars and 1 BL Lac
within the $\mu_B$=26$^m$/$\Box\arcsec$ isophote.

The sample selection is heterogeneous, because of the non-uniform
coverage of different surveys that provided the photometric data.
Therefore, it is difficult to estimate the selection bias and
completeness of our search (see Sect. 4.2).

Thirteen spectra came from a campaign to confirm optical and X-ray
selected high mass binaries in M\,31 \citep{2014MNRAS.443.2499W}; 12
of them were identified as quasars and one was a M star (No. 14 in
Table\,\ref{tab:obslog}).

\begin{figure}
\includegraphics[width=8.9cm]{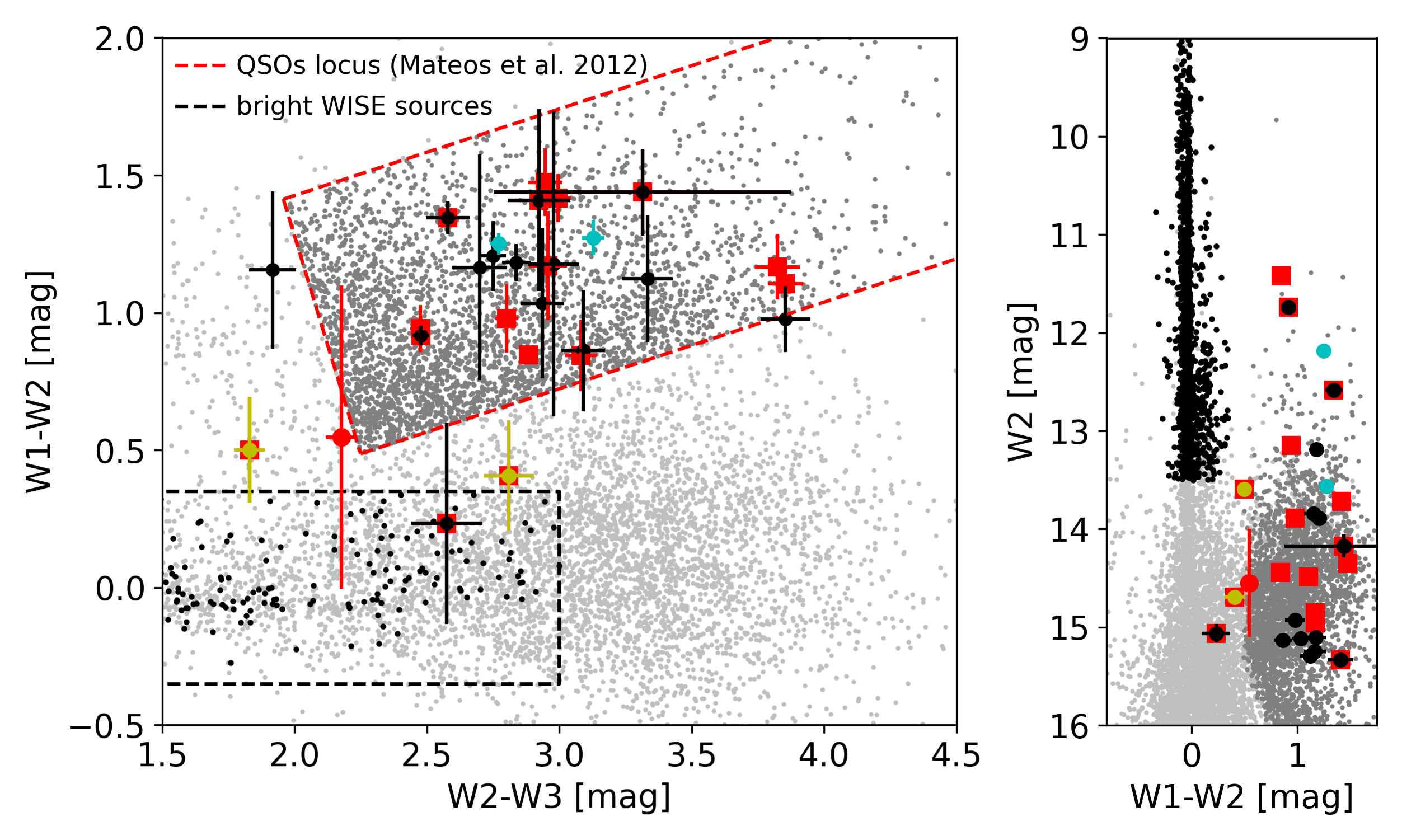}
\includegraphics[width=8.9cm]{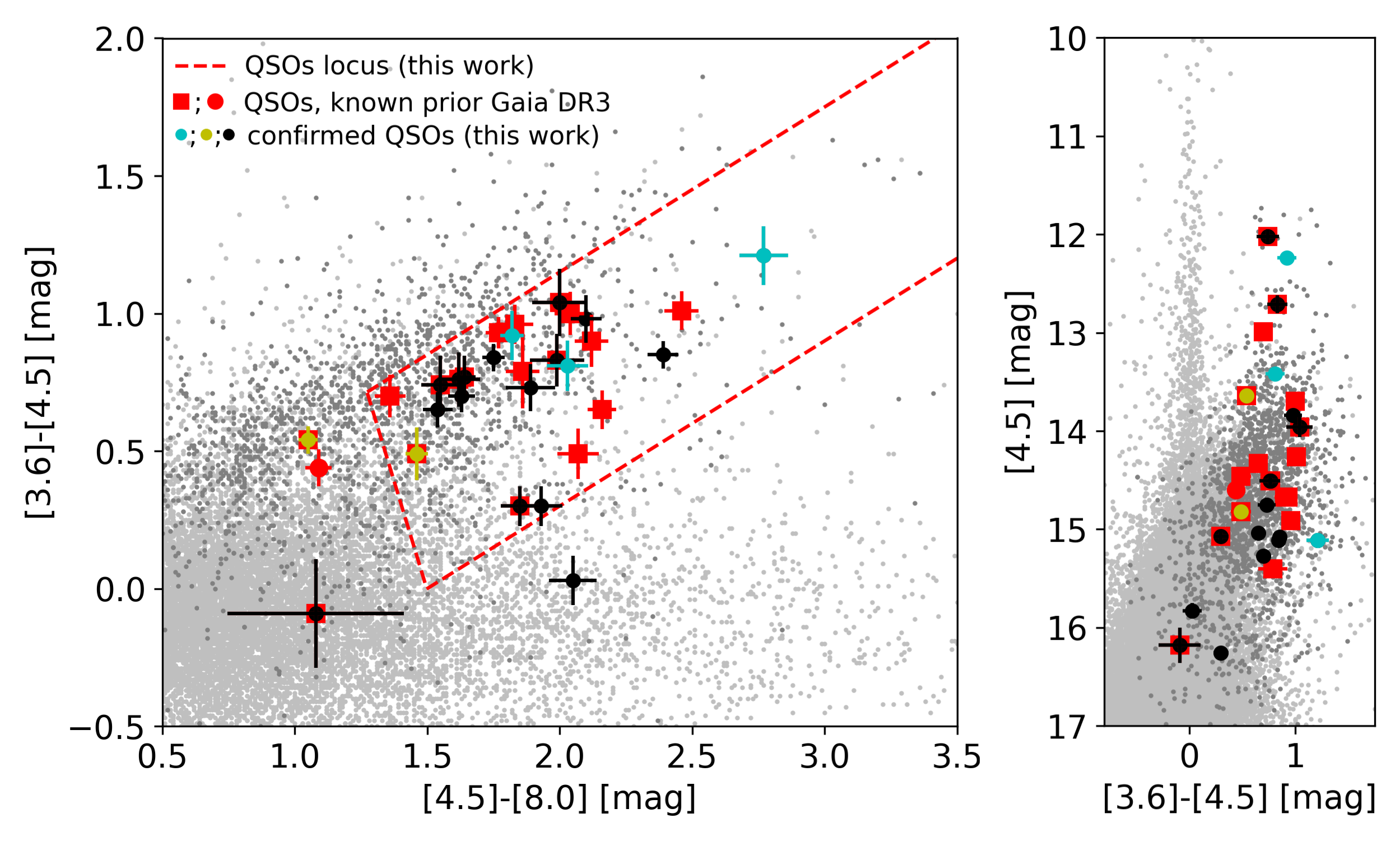}
\includegraphics[width=8.99cm]{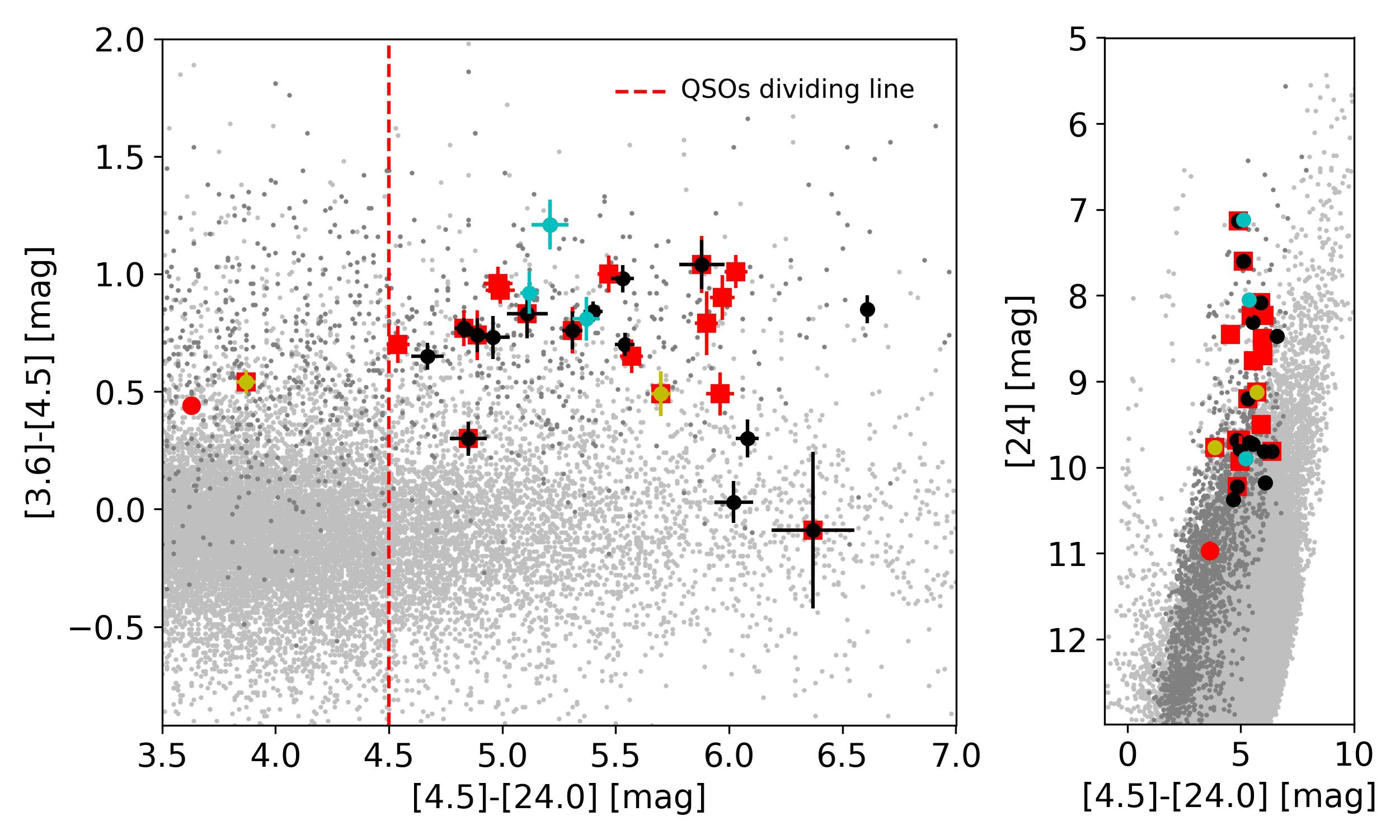}
\caption{\small Illustration of our MIR color selection (see
Sec.\,\ref{sec:selection} for details).
{\it Top}: {\it WISE} color--color (left) and color-magnitude
diagram (right). The symbols in all panels are the same as in
Fig.\,\ref{fig:qso_before_GDR3.png}.
Small gray dots show the entire WISE photometry within the {\it
Spitzer} footprint. Small black solid dots
are the sample of bright {\it WISE} sources used to adjust the
coordinates (see Appx.\,\ref{app:mid_ir_coords} for details). The
red dashed lines show our adopted quasar locus defined in
Eqs.\,\ref{eq:MIR_selection} after \citet{2012MNRAS.426.3271M}.
The sources falling inside are plotted with dark gray in all panels.
{\it Middle and bottom}: Similar diagrams, but with {\it Spitzer}
3.6, 4.6 and 8 or 24 $\mu$m bands. The dashed vertical line in the
bottom plot shows our additional color constraint. Summarizing,
our targets for follow up spectroscopy were selected among dark
gray points that meet this additional criterion.
}\label{fig:quasar_ccd_cmd_W}
\end{figure}

Next, we expanded our search to the mid-IR. Initially, its
footprint was limited to the Panchromatic Hubble Andromeda
Treasury survey \citep[PHAT;][]{2014ApJS..215....9W} where deep
and high angular resolution imaging allowed to identify isolated
bright candidates that meet the mid-IR criteria of
\citet{2012MNRAS.426.3271M}, based on colors from the {\it
Wide-field Infrared Survey Explorer}
\citep[WISE;][]{2010AJ....140.1868W}. This is shown in
Fig.\,\ref{fig:quasar_ccd_cmd_W} (top panel). We obtained
spectra of 4 candidates from this selection.

We further expanded the target selection to include the Local
Group Galaxy survey \citep[LGGS;][]{2016AJ....152...62M} combining
it with mid-IR {\it Spitzer} photometry of \citet{2017ApJS..228....5K}.
Following \citet{2012MNRAS.426.3271M}, we defined a locus
(Fig.\,\ref{fig:quasar_ccd_cmd_W}, middle) of known quasars in the
{\it Spitzer} colors space as follows:
\begin{equation}\label{eq:MIR_selection}
\begin{array}{l}
([3.6]-[4.5])\geq-3.17\times([4.5]-[8.0])+4.75\\
([3.6]-[4.5])\geq0.6\times([4.5]-[8.0])-0.90\\
([3.6]-[4.5])\leq0.6\times([4.5]-[8.0])-0.05
\end{array}
\end{equation}

This is an intentionally conservative definition that excludes
some known quasars, but it
helps to reduce the contamination by red stars. In addition,
we required that ([4.5]$-$[24])$\geq$4.5\,mag and that the
sources are isolated -- the degree of crowding index defined
by \citet{2016AJ....152...62M} is $<$5\%.
Nineteen ($\sim$85\%) of the 22 known quasars fall within
the locus defined by these criteria; the three remaining
sources are nearby AGNs where the nuclei may be contaminated
by the emission from the galaxy disks
(Fig.\,\ref{fig:quasar_ccd_cmd_W}, bottom).
Even these conservative criteria yielded hundreds of potential
QSOs, but only a handful of them were bright enough in the
optical for follow up spectroscopy: 10 have $R$$\leq$19.2\,mag,
32 with 19.2$<$$R$$\leq$20.0\,mag, and 62 with
20.0$<$$R$$\leq$20.5\,mag, giving us in total 104 LGGS objects
within a searching radius of 1.5$\arcsec$. We
obtained spectra of only 8 of these.

Seven quasars from Yuhan
Yao\footnote{\url{https://speakerdeck.com/yaoyuhan/quasars-behind-m31-from-ptf-survey}}
were reidentified among known X-ray \citep{2015PASA...32...10F}
and reobserved producing homogeneous redshifts for 6 of them.
The remaining object (No. 4 in Table\,\ref{tab:obslog}) was
classified as an X-ray binary.

The complete list of 32 observed candidates is given in
Table\,\ref{tab:obslog}.
In the process of matching the optical and mid-IR surveys, we
discovered some systematic differences in the coordinates of
\citet{2017ApJS..228....5K}. The applied corrections with
respect to WISE catalog are described in
Appendix\,\ref{app:mid_ir_coords}.

\section{Observations and data reduction}\label{sec:obs}

The spectroscopic confirmation was obtained by low-resolution optical
spectroscopy that allows to recognize quasars from the broad emission
lines in their spectra.
Part of the spectra were obtained with the Dual Imaging
Spectrograph\footnote{\url{https://www.apo.nmsu.edu/arc35m/Instruments/DIS/}}
(DIS) at the 3.5m telescope at the Apache Point Observatory (marked with
A in column Telescope in Table\,\ref{tab:obslog}), with the B400 and R300
gratings in the blue ($\lambda$$\approx$3425--5900\,\AA) and red
($\lambda$$\approx$5180--9860\,\AA) channel, respectively. The 1.5\,arcsec
slit was used, delivering resolving power of R$\approx$700 and 850 at the
centers of the wavelength ranges for the blue and red arms, respectively.

The remaining spectra were obtained at the 6m telescope at the Special
Astrophysical Observatory (SAO) of the Russian Academy of Sciences (marked
with B in column Telescope in Table\,\ref{tab:obslog}) with
SCORPIO-1\footnote{\url{https://www.sao.ru/hq/lsfvo/devices/scorpio/scorpio.html}}
\citep[grisms VPHG1200B and VPHG1200R, spectral ranges and resolutions
are $\lambda$$\approx$3600--5400\,\AA\ and R$\approx$800,
$\lambda$$\approx$5700--7400\,\AA\ R$\approx$1300,
respectively; 1.2\,arcsec wide slit;][]{2005AstL...31..194A} and
SCORPIO-2\footnote{\url{https://www.sao.ru/hq/lsfvo/devices/scorpio-2/index.html}}
\citep[grism VPHG1200@540, spectral range $\lambda$$\approx$3650--7250\,\AA,
resolution R$\approx$1000; 1.0\,arcsec wide slit;][]{2011BaltA..20..363A}.
Table\,\ref{tab:obslog} shows the log of our observations.

\begin{table*}[!ht]
\caption{Observing log. The columns contain: target identification,
position, Gaia G band magnitude, telescope (A -- Apache Point 3.5m,
B -- BTA 6m), UT at the start of the first exposure, number and integration
time for the individual exposures (occasionally, the last exposure is
shorter because the observations had to be interrupted due to poor weather
and there may be a different number of exposures for the blue B and the
red R channels), cross-identification with Table\,\ref{tab:redshifts},
selection criteria and comments. The astrometry and photometry is from
Gaia\,DR3 \citep{2023A&A...674A...1G, 2023A&A...674A...41A},
except for the object No. 5, marked with ``a''; for it the data are from
Pan-STARRS Survey DR1 \citep{2016arXiv161205560C}.
Asterisks indicate 16 quasars whose spectra are published for the first time.}
\label{tab:obslog}
{\small
%\begin{tabular}{@{}l@{ }l@{ }l@{ }l@{ }l@{ }l@{ } }l@{ }@{ }l@{ }l{}}
\begin{tabular}{@{}lcc@{ }c@{ }c@{ }c@{ }c@{ }c@{ }p{4.2cm}@{}}
\hline
No  & Gaia & RA DEC  & G,    &~Tel.~~& UT start     & N$_{\rm exp}$$\times$ & No. in~           &~Selection criteria; comments \\
    & EDR3 & (J2000) &~~mag~~&       & 20yy-mm-dd   & int.\,                & Table               &          \\
    &      &         &       &       & Thh:mm       & time,\,s              & \protect{\ref{tab:redshifts}} &          \\
\hline
1*  & 381231021002898048     & 00:39:21.921 41:20:31.31 & 20.52     & B & 21-11-03T21:18~ & 6$\times$900     &  15 & Opt. colors \& {\it PTF} variability\\
2*  & 381230200663395712     & 00:39:28.105 41:19:13.09 & 20.77     & B & 21-11-03T23:04~ & 4$\times$900     &  16 & Opt. colors \& {\it PTF} variability\\
3*  & 381223637953333376     & 00:39:34.595 41:07:38.52 & 19.88     & B & 21-11-04T19:34~ & 3$\times$900     &  18 & Opt. colors \& {\it PTF} variability\\
4*  & 381251297542795904     & 00:39:47.423 41:31:47.53 & 20.31     & B & 21-10-12T17:17~ & 3$\times$900     &  21 & Opt. colors \& {\it PTF} variability\\
5   &~156600101021498792$^a$~& 00:40:24.514 40:30:25.12 &~20.00$^a$~& B & 21-12-12T20:37~ & 8$\times$600     &     & Opt. colors \& {\it PTF} variability; not QSO, X-ray binary? \\
6*  & 381129011231416704     & 00:40:29.727 40:37:05.68 & 18.98     & A & 07-11-03T01:34~ & 3$\times$2700    &  27 & X-ray \& {\it LGGS}\\
    &                        &                          &           & A & 07-11-09T02:07~ & 3$\times$2700    &     & \\
7   & 381159003004920192     & 00:40:45.721 40:51:34.18 & 19.24     & A & 20-01-02T04:55~ & 2$\times$1800    &     & {\it Spitzer} colors \& {\it LGGS}; too low S/N to interpret \\
8   & 381161584267072896     & 00:41:26.217 40:53:26.55 & 19.95     & A & 07-10-10T04:38~ & 3$\times$2700    &  41 & X-ray \& {\it LGGS}\\
9   & 381164710999391104     & 00:41:37.918 41:01:07.82 & 19.40     & A & 07-08-10T08:55~ & 4$\times$1800    &  45 & X-ray \& {\it LGGS}\\
10  & 381276410213544064     & 00:41:41.408 41:19:16.85 & 18.57     & A & 07-10-09T02:05~ & 3$\times$2700    &  46 & X-ray \& {\it LGGS}\\
11* & 381266244030810880     & 00:42:15.487 41:20:31.52 & 20.54     & A & 10-10-30T02:46~ & 6$\times$2700    &  57 & X-ray \& {\it LGGS}\\
12* & 369148728241389056     & 00:42:35.002 40:48:39.20 & 18.81     & A & 07-08-11T10:55~ & 3$\times$1200    &  62 & X-ray \& {\it LGGS}\\
    &                        &                          &           & A & 10-10-30T01:34~ & 3$\times$1500    &     & \\
13  & 369257652908605824     & 00:43:12.746 41:16:25.49 & 18.58     & A & 18-10-07T02:27~ & 2$\times$1800    &     & UV excess in {\it LGGS} \& {\it PHAT}; not QSO, star forming region or galaxy? \\
14  & 369255522604831232     & 00:43:53.335 41:16:55.90 & 19.51     & A & 07-08-08T07:59~ & 2$\times$1800    &     & X-ray \& {\it LGGS}; not QSO, early M star \\
15  & 369286102772174208     & 00:44:10.843 41:33:01.15 & 19.22     & A & 19-01-08T05:16~ & 2$\times$1500,   &     & {\it Spitzer} colors \& {\it LGGS}; flat feature-\\
    &                        &                          &           & A &                 & 1$\times$602     &     & less spectrum \\
%   &                        &                          &           & A & 19-09-05T07:18~ & 3$\times$1200    &     & 2 emission lines not interpretable \\
16  & 369281288111420288     & 00:44:28.716 41:25:45.62 & 18.96     & A & 19-01-06T04:03~ & 2$\times$1500,   &     & {\it Spitzer} colors \& {\it LGGS}; not quasar,\\
    &                        &                          &           & A &                 & 1$\times$1034    &     & K-type star or globular cluster, at the redshift of M\,31\\
    &                        &                          &           & A & 19-01-08T04:17~ & 2$\times$1200    &     &  \\
    &                        &                          &           & A & 19-09-05T08:25~ & 3$\times$1500    &     &  \\
    &                        &                          &           & B & 22-10-25T17:33~ & 3$\times$900\,B  &     & \\
    &                        &                          &           & B & 22-10-25T18:25~ & 2$\times$900\,R  &     & \\
17  & 369281086247278080     & 00:44:31.624 41:24:10.69 & 19.86     & A & 19-10-27T03:35~ & 6$\times$1800    &     & {\it Spitzer} colors \& {\it LGGS}; not quasar, wide emission lines object \\ %M31 eclipsing binary (Vilardell+, 2006) \\
18* & 369185428738695808     & 00:44:41.590 40:46:43.46 & 19.10     & B & 21-10-02T22:11~ & 3$\times$900\,B  & 79 & Opt. colors \& {\it PTF} variability\\
    &                        &                          &           & B & 21-10-02T22:59~ & 2$\times$900\,R  &     & \\
19  & 381304653920639488     & 00:44:45.117 41:47:20.52 & 19.31     & A & 19-10-27T01:13~ & 4$\times$1800    &     & {\it Spitzer} colors + {\it LGGS}; too low S/N to interpret \\
20* & 369267681654714112     & 00:44:48.307 41:16:18.23 & 18.43     & A & 19-01-06T01:57~ & 3$\times$900     & 82 & {\it Spitzer} colors + {\it LGGS}\\
    &                        &                          &           & A & 20-01-02T01:57~ & 2$\times$1800    &     & \\
21* & 369189547608980224     & 00:44:57.601 40:59:13.85 & 20.47     & B & 21-12-11T15:20~ & 4$\times$600     & 83 & Opt. colors \& {\it PTF} variability\\
22* & 369268712447126272     & 00:44:57.938 41:23:43.72 & 19.67     & A & 07-11-09T03:45~ & 2$\times$1800    & 84 & X-ray \& {\it LGGS}\\
23  & 369288714109887744     & 00:45:27.311 41:32:54.06 & 20.39     & A & 18-10-07T08:03~ & 4$\times$1800    & 92 & UV excess in {\it LGGS} \& {\it PHAT}\\
24* & 369276756923364224     & 00:45:28.249 41:29:43.92 & 19.53     & A & 18-10-07T05:08~ & 4$\times$1800    & 93 & UV excess in {\it LGGS} \& {\it PHAT}\\
25  & 369270469091280000     & 00:45:43.426 41:20:30.86 & 19.27     & A & 07-07-16T08:29~ &~3$\times$1800\,B~& 95 & X-ray \& {\it LGGS}\\
    &                        &                          &           & A &                 &~4$\times$1500\,R~&     & \\
26  & 375332454650675328     & 00:46:32.809 42:13:14.25 & 19.10     & A & 18-10-07T04:06~ & 3$\times$900     &     & UV excess in {\it LGGS} \& {\it PHAT}; featureless spectra, no emission lines \\
27  & 375328709439174400     & 00:46:55.515 42:20:50.09 & 17.65     & A & 10-10-30T05:12  & 3$\times$900     & 108 & X-ray \& {\it LGGS}\\
28* & 375328395906539392     & 00:47:02.870 42:18:37.19 & 19.90     & A & 10-10-10T07:37~ & 8$\times$836     & 109 & X-ray \& {\it LGGS}\\
29* & 375328876942905472     & 00:47:13.476 42:20:47.66 & 19.92     & A & 07-07-24T08:05~ & 4$\times$1800    & 111 & X-ray \& {\it LGGS}\\
30* & 375423881617811328     & 00:47:39.120 42:26:17.76 & 18.20     & A & 19-01-06T02:54~ & 3$\times$1200    & 115   & {\it Spitzer} colors \& {\it LGGS}\\
31  & 375322314232658048     & 00:47:42.847 42:10:16.93 & 19.74     & A & 08-10-26T01:29~ & 3$\times$2700    & 116 & X-ray \& {\it LGGS}\\
32* & 375418590218050304     & 00:47:48.323 42:19:34.31 & 19.13     & A & 19-01-08T03:07~ & 3$\times$1200    & 118 & {\it Spitzer} colors \& {\it LGGS}\\
    &                        &                          &           & A & 20-01-02T03:10~ & 3$\times$1800    &     & \\
\hline
\end{tabular}
}
\end{table*}

\begin{figure*}
\centering
\includegraphics[width=9.15cm]{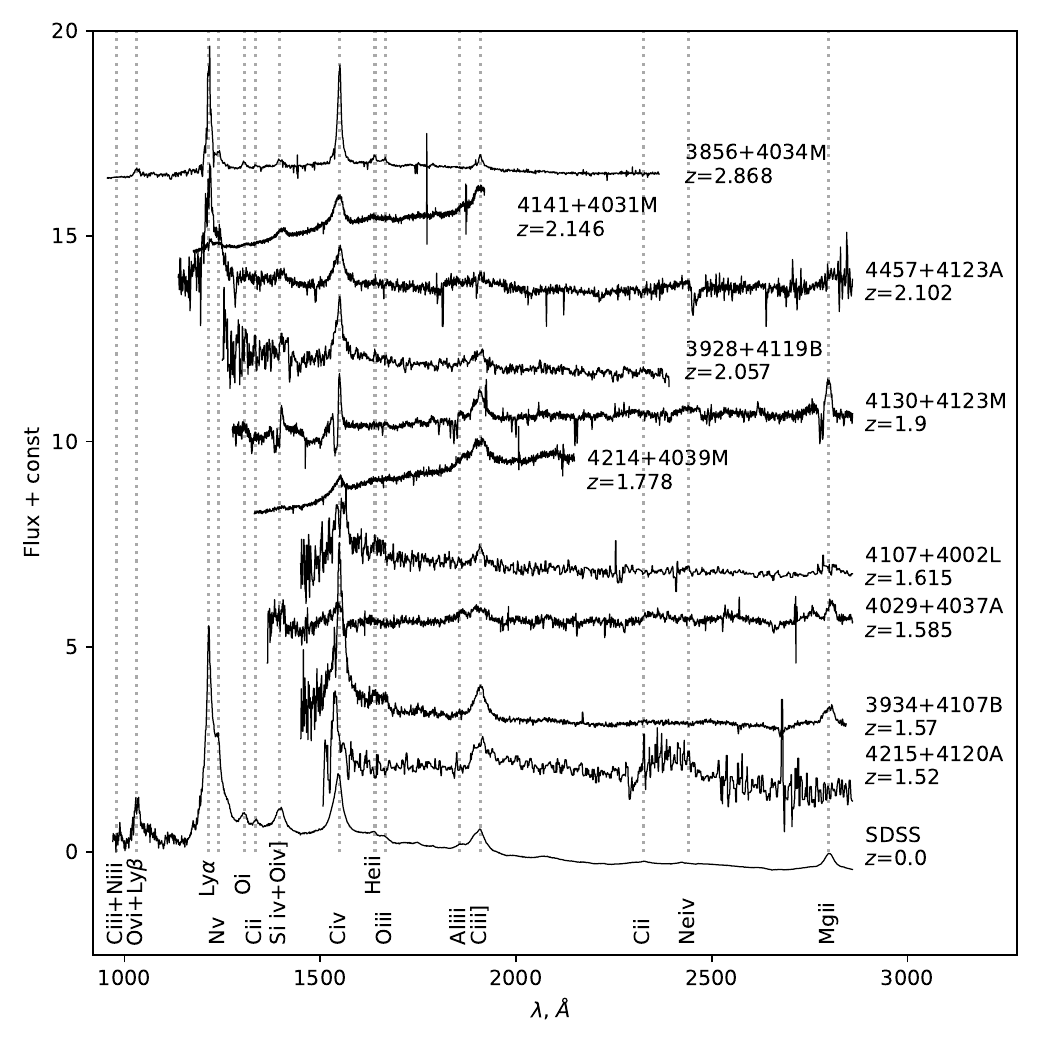}
\includegraphics[width=9.15cm]{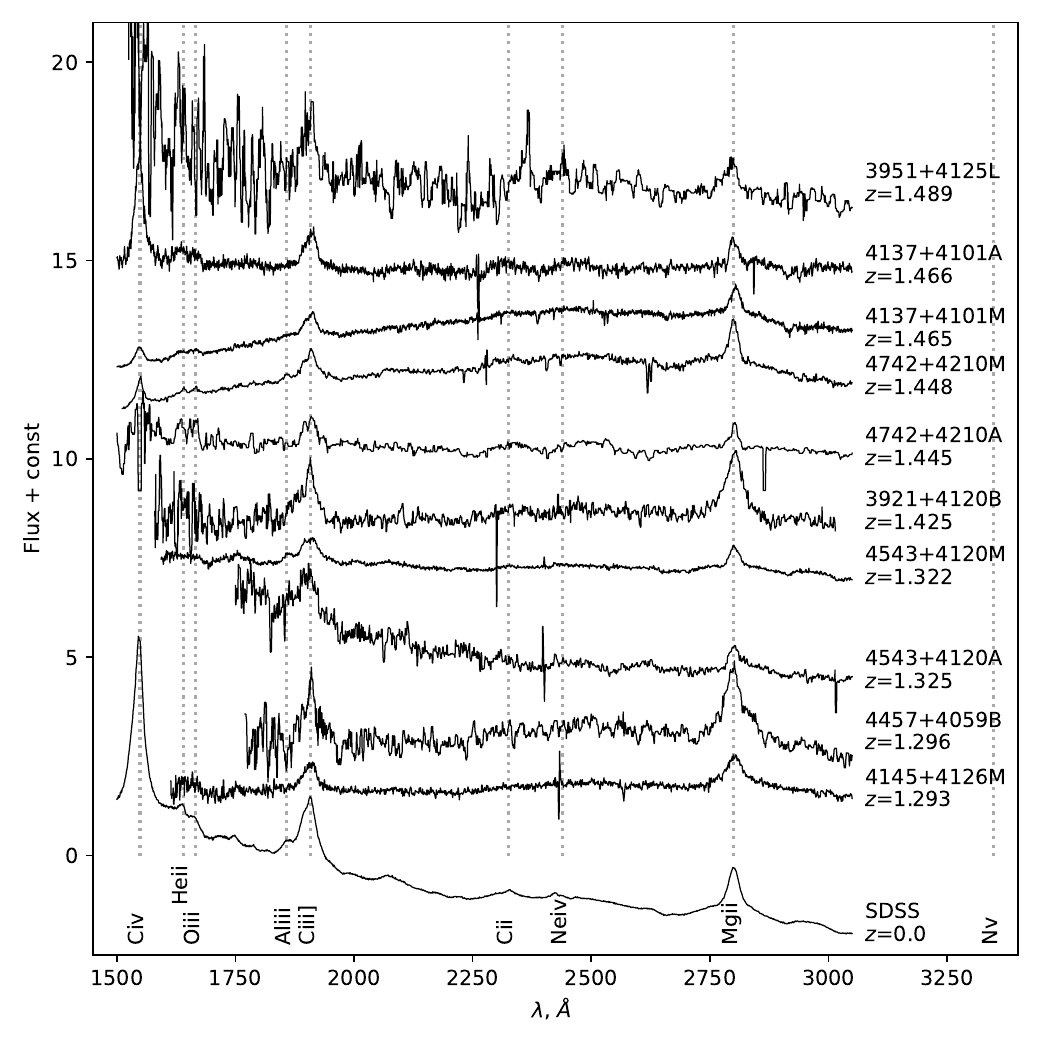}
\includegraphics[width=9.15cm]{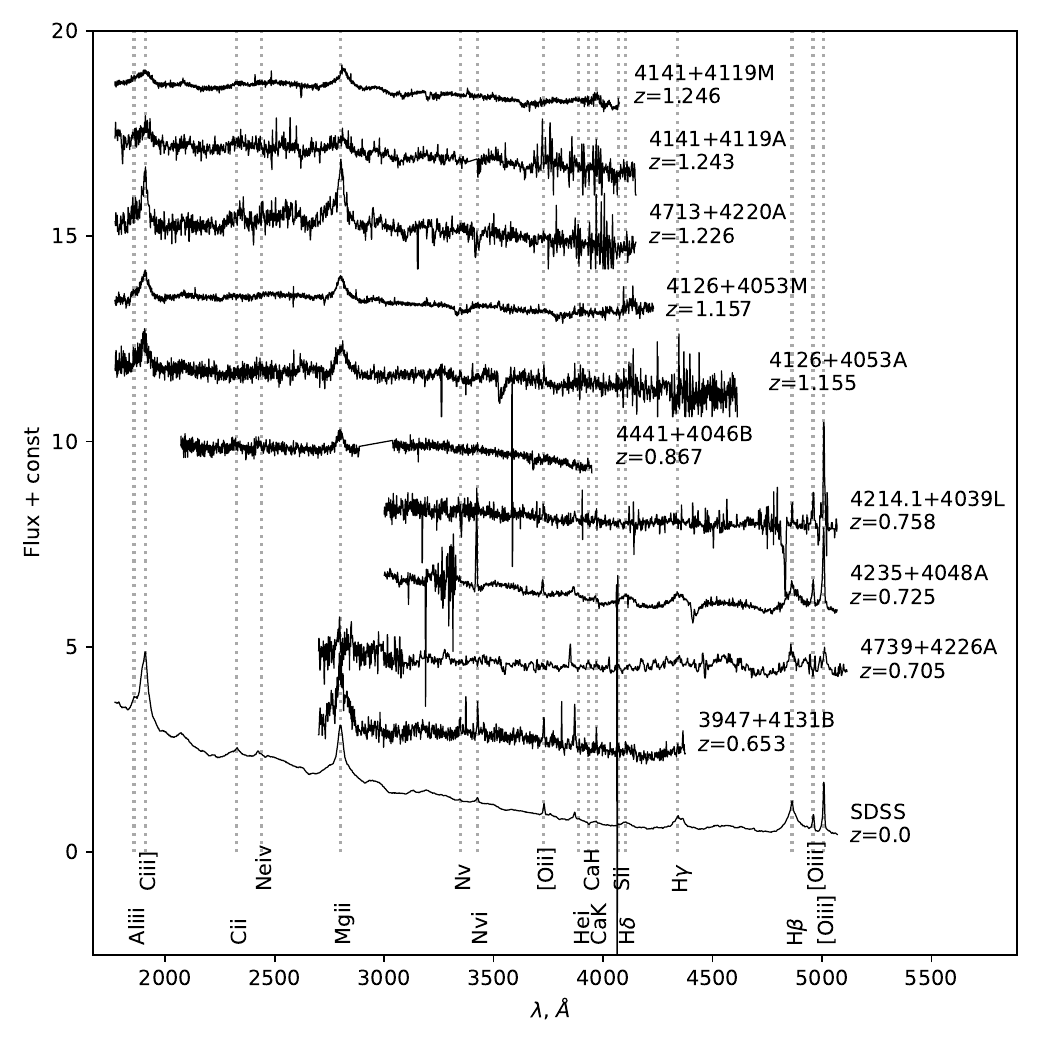}
\includegraphics[width=9.15cm]{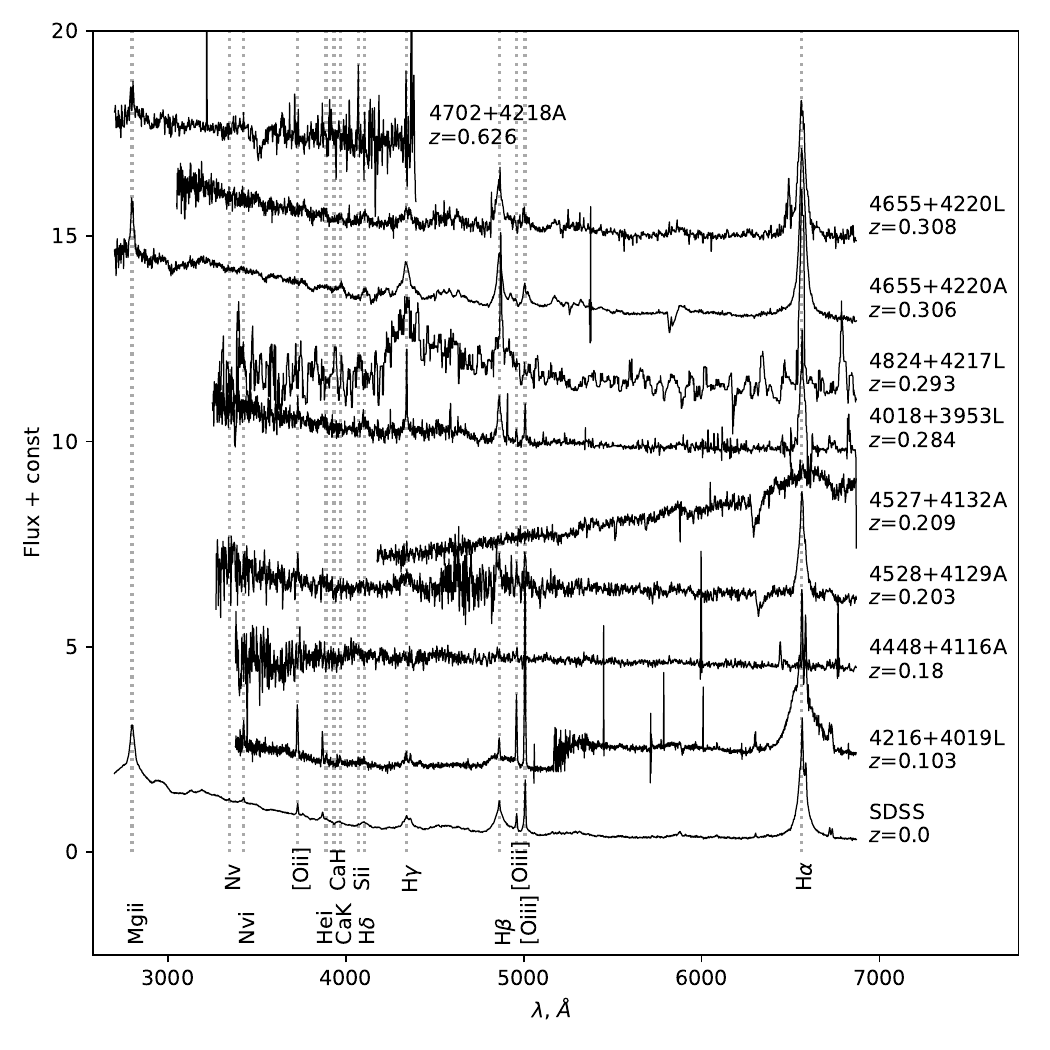}
\caption{Spectra of the observed objects sorted by redshift, shown
at rest frame wavelength. Some more prominent features are marked
with vertical dashed lines and are labeled. The SDSS composite
quasar spectrum \citep{2001AJ....122..549V} is also plotted at the
bottom of each panel.}\label{fig:quasar_spectra}
\end{figure*}

The data were reduced with Image Reduction and Analysis
Facility\footnote{IRAF is distributed by the National Optical
Astronomy
Observatory, which is operated by the Association of Universities
for Research in Astronomy under a cooperative agreement with the
National Science Foundation.}
\citep[IRAF;][]{1993ASPC...52..173T,1986SPIE..627..733T}
and ESO Munich Image Data Analysis
System\footnote{\url{https://www.eso.org/sci/software/esomidas/}}
\citep[MIDAS;][]{1988igbo.conf..431B,1992ASPC...25..115W}.
The usual steps for long slit data processing were performed:
bias/dark subtraction, flat fielding and wavelength calibration.
The 1-dimensional spectra were extracted within 0.8--1.2\,arcsec
wide apertures, approximately matching the seeing during the
observations for the DIS data, or using the optimal extraction
code SPEXTRA, specifically designed for crowded fields
\citep{2017AstBu..72..486S}, for the SCORPIO data. The spectra
were flux calibrated with spectrophotometric standards from
\citet{1990AJ.....99.1621O} and
\citet{1992PASP..104..533H,1994PASP..106..566H},
but given the non-photometric conditions during most of the
observations, it is relative, so we only recovered
the correct shape of the spectra, not their absolute flux.
The quasar spectra are plotted in Fig.\,\ref{fig:quasar_spectra}.

Broad-band $grizy$ photometry for the quasars behind M\,31 and
in the surrounding field was obtained from Pan-STARRS Survey DR1
\citep{2016arXiv161205560C}.

\section{Quasar sample}\label{sec:quasar_sample}

The presence of broad lines in our spectra indicates
that 23 (yellow, cyan and black solid dots in
Fig.\,\ref{fig:qso_before_GDR3.png}) out of the 32 observed objects are
quasars; J004029.727+403705.68 and J004215.489+412031.52 (No. 6
and 11 in Table\,\ref{tab:obslog}, respectively) are new and the
rest are confirmations, although for 16 of these our spectra are
the first to be published;
the 9 non-quasars are stars, galaxies or their nature can not be
reliably identified because of the low signal-to-noise of the
spectra. Among 13 X-ray selected candidates, 12 turn out to be
quasars; the UV excess in LGGS-PHAT selected 4 candidates yielded
2 quasars; the SPITZER-LGGS selected 8 candidates also yielded
3 quasars; 6 of 7 Yuhan Yao objects appear to be quasars.
Analyzing 17 more archival spectra, in addition to our data,
we uniformly measured redshifts for in total of 34 unique quasars
based on 40 spectra in total.

The objects with measured redshift in this work together with
the literature QSOs with spectroscopically derived redshift form
a sample of 125 QSOs, located within M\,31 isophote
$\mu_B$=26$^m$/$\Box\arcsec$. They are listed in
Table\,\ref{tab:redshifts}, together with the lines used for the
redshift measurements and the distances from the M\,31 center.
For the center of M\,31, we adopted the coordinates of its extended
core J004244.33+411607.5 \citep{2006AJ....131.1163S} located at
1.7$\arcsec$ from the center defined in RC3 \citep{1991rc3..book.....D}.
All distances $\rho$ within the rectified plane of the galaxy are
in fractions, normalized to $\rho_{25}$, which is the half of
the major isophotal diameter D25=190.5$\arcmin$
measured at surface brightness level $\mu_B$=25$^m$/$\Box\arcsec$.
They are calculated, assuming position angle PA=35\,deg, determined
as the ratio of the major-to-minor isophotal diameters R25=3.0903
\citep{1991rc3..book.....D} and taking into account an intrinsic
oblateness of 0.14, typical for a Sb galaxy like M\,31. Thus, the
disk plane is inclined to the line of sight at 17.1\,deg.

To outline the area of interest in this work, we tentatively
adopted a larger major isophotal diameter of 225$\arcmin$ as
measured approximately at surface brightness level
$\mu_B$=26$^m$/$\Box\arcsec$ by \citet{1987A&AS...69..311W}, the
same axis ratio and PA as in RC3. Note that the Holmberg diameter
of 200$\arcmin$ \citep{1973UGC...C...0000N} falls just between
the previous two values, but axis ratio of 2.5 leads to a smaller
inclination of 12.1\,deg.

\begin{figure*}[!ht]
\centering
\includegraphics[width=18.65cm]{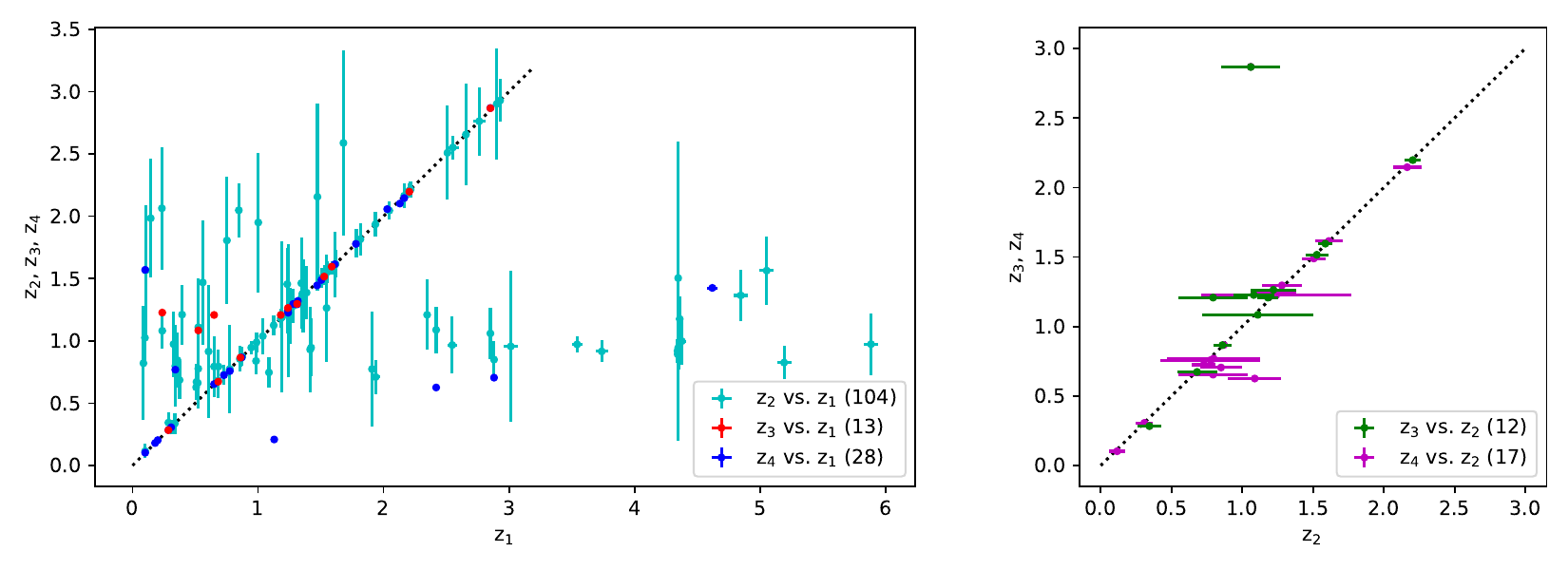}
% \resizebox{\hsize}{!}{\includegraphics{redshifts_cross_check_b.png}}
\caption{Cross-comparison of redshifts from different sources:
$z_1$ -- \citet{2023A&A...674A...1G};
$z_2$ -- \citet{2024ApJ...964...69S};
$z_3$ -- \citet{2023ApJ...944....1D};
$z_4$ -- all other sources (listed in Table\,\ref{tab:redshifts}),
including this work redshift preference when available. Left: $z_2$,
$z_3$ and $z_4$ versus $z_1$. Right: $z_3$ and $z_4$ versus $z_2$.
The published measurement
errors are shown, except for \citet{2023ApJ...944....1D} who do not
list them -- in this case for plotting purposes we adopted errors
equal to 0.005\,dex. The bracketed numbers in
the legends show how many objects constitute the overlapping subset
between each pair of samples. The black dotted line has a slope of
one.
%{\bf Add that many outliers have Pqso=1 so are not reliable!}
}
\label{fig:redshifts_comp}
\end{figure*}

\subsection{Redshift measurements}\label{sec:analysis}

Quasars are easy to identify from the broad (as wide as thousands
of km\,s$^{-1}$) lines, like C{\sc iv} 1548, C{\sc iii}] 1909,
Mg{\sc ii} 2799 and others, emitted by the gas in the vicinity of
the central supermassive black holes
\citep[BH;][]{1997ApJ...479..642B,1997iagn.book.....P}.

We measured their redshift following closely the procedures of
\citet{2016A&A...588A..93I} and \citet{2019AJ....157..227M}. First,
we identified the broad lines comparing the spectrum of each object
with the composite SDSS quasar spectrum of \citet{2001AJ....122..549V}.
Then, we measured the observed wavelengths of the lines fitting
Gaussians to them with the IRAF task {\it splot} and calculated
the redshift of each line. Blended lines were omitted. Lines
affected by strong intervening absorption were either omitted, or
if possible, Gaussians were fitted only to their cores. The
final redshift of an object was determined averaging the
redshifts of all emission lines. The redshift uncertainty was
the r.m.s. of the individual lines' redshifts; if only two lines
were available, we tentatively adopted as an error the difference
between their redshifts.

\subsection{Comparison with literature redshifts}

In general, our redshift measurements $z_4$ shown with dots in
Fig.\,\ref{fig:redshifts_comp} (right), agree well with the
literature, with a few exceptions. For example, we measured
$z$=0.626$\pm$0.011 for Gaia\,DR3 375328395906539392 (Nos. 28 in
Table\,\ref{tab:obslog} and 109 in Table\,\ref{tab:redshifts}),
while \citet{2023A&A...674A...1G} and \citet{2024ApJ...964...69S}
reported the vastly different
$z$=2.419$\pm$0.016 and $z$=1.088$\pm$0.188, respectively. We
identified the single emission line in its spectrum as Mg{\sc ii}\,2799,
because no other prominent lines appeared within the wavelength
coverage of our APO spectrum, and they should have, if the redshift
was different.

For Gaia\,DR3\,375423881617811328 (Nos. 30 in Table\,\ref{tab:obslog}
and 115 in Table\,\ref{tab:redshifts}), we measured
$z$=0.705$\pm$0.001, agreeing at $\sim$1$\sigma$ level with
$z$=0.850$\pm$0.147 of \citet{2024ApJ...964...69S}, but very
different from $z$=2.880$\pm$0.026 of \citet{2023A&A...674A...1G}.
Our measurement is based on three lines, making the random
agreement unlikely.

Gaia\,DR3\,369288714109887744 (Nos. 23 in Table\,\ref{tab:obslog} and
92 in Table\,\ref{tab:redshifts}) is another case of disagreement.
This is a rediscovered quasar at z=0.209$\pm$0.001 found earlier by
\citet{2017ApJ...850...86D} at z$\sim$0.215. A third measurement of
z=1.129$\pm$0.008 \citep{2023A&A...674A...1G} is considered
unreliable.

The largest redshift discrepancy is associated with
Gaia\,DR3 381231021002898048 (Nos. 1 in Table\,\ref{tab:obslog} and
15 in Table\,\ref{tab:redshifts}). Our higher quality SAO spectrum
yields 1.425$\pm$0.005, very different from z=4.619$\pm$0.042 of
\citet{2023A&A...674A...1G}. Notably, this work gives some extremely
high redshifts -- with respect to \citet{2024ApJ...964...69S} -- for
other 21 qausars.

For two objects, we measured redshifts higher than the values of
\citet{2023A&A...674A...1G}. For DR3 375418590218050304 (Nos. 32 in
Table\,\ref{tab:obslog} and 118 in Table\,\ref{tab:redshifts}), they
reported $z$=0.341$\pm$0.003, while we measured $z$=0.769$\pm$0.002,
very similar to $z$=0.798$\pm$0.328 of \citet{2024ApJ...964...69S}.
The decisive argument for our higher redshift is the identification
of Mg{\sc ii}\,2799 in the blue arm of the APO spectrograph. The two
arms are calibrated independently, in effect they are two instruments.

For Gaia\,DR3\,381223637953333376 (Nos. 3 in Table\,\ref{tab:obslog}
and 18 in Table\,\ref{tab:redshifts}), we measured $z$=1.570$\pm$0.001.
\citet{2023A&A...674A...1G} place it at $z$=0.101$\pm$0.008 --
inconsistent with our excellent quality SAO spectrum that exhibits
multiple emission lines.

Not surprisingly, somewhat better agreement exists between $z_3$
\citep{2023ApJ...944....1D} and $z_2$ \citep[] {2024ApJ...964...69S},
and between our redshifts $z_3$ and $z_2$ \citep[] {2024ApJ...964...69S}
as seen in Fig.\,\ref{fig:redshifts_comp}, right.
The only inconsistent case here is Gaia\,DR3\,381137287633113216 (No.
7 in Table\,\ref{tab:redshifts}): \citet{2024ApJ...964...69S}
reported $z$=1.060$\pm$0.209, very different from our value of
$z$=2.868$\pm$0.003. Our value agrees with three other measurements
from the literature.

The blazar Gaia\,DR3\,375378320605062144 \citep{2017ApJ...851..135P}
was excluded from our redshift and reddening analyses, because of
the contradicting $z$ measurements in the literature and because it
may not exhibit typical quasar colors.

Some objects were misidentified as quasars in the literature.
\citet{2019RAA....19...29L} reported that LAMOST\_10.952+40.80076
(Gaia\,DR3\,369228137893480064) is a QSO at $z$=0.262. However, a
careful inspection of the LAMOST DR5 spectrum
(spec-55859-M5901\_sp04-015) revealed prominent CaT and other
stellar features typical for K7-M0 stars. Another example is Gaia
DR3 381200372113648128, listed as J003905.66+410456.6 in LAMOST
DR2\&3 \citep{2018AJ....155..189D} and LAMOST\_9.774+41.08240 in
LAMOST DR4 \citep{2019RAA....19...29L}. We did not find convincing
confirmation of the $z$$\sim$0.8 derived in both these papers from
the LAMOST spectra. Instead, we identified a broad feature
corresponding to H$\alpha$ at the M\,31 rest frame and we suspected
that it has been misinterpreted as C{\sc iv}\,1548 by
\citet{2023A&A...674A...1G} who reported $z$=3.190$\pm$0.027.

Our selection identified 9 more QSO candidates within the
$\mu_B$=26$^m$/$\Box\arcsec$ isophote and we observed them (Nos.
5, 7, 13--17, 19, 26 in Table\,\ref{tab:obslog}), but the spectra
did not provide evidences for a quasar nature. This was the
case with object No. 16, located in a prominent dust lane.
\citet{2015ApJ...802..127J} reported $\sim$10\% probability that
it is a galaxy. However, our spectrum lacks broad emission lines
and shows absorptions making it a likely red star or a globular
cluster in M\,31. These nine and the two LAMOST objects mentioned
above were omitted from Table\,\ref{tab:redshifts}.

The heterogeneous nature of the selection and the lack of
comparison quasars outside of the M\,31 disk, confirmed by the
same means and following the same strategy makes it difficult
to estimate the completeness of this sample. The only exception,
and the largest contribution to our sample -- bringing in about
2/3 of the quasars -- is the all-sky Quaia survey. We compared
the surface density of secure quasars, according to the
classification of \citet[][see their Class, reproduced here in
Table\,\ref{tab:redshifts}]{2023A&A...674A...1G}. There is an
10-30\% excess of such objects inside of the
$\mu_B$=26$^m$/$\Box\arcsec$ isophote, where the exact percentage
depends on the adopted outside locus and magnitude limits.
Therefore, it is not the incompleteness but rather the
contamination that may be an issue for quasar searches,
underlying the need for reliable spectroscopic confirmation like
the follow up we report here.

\subsection{Reliability of redshift measurements}

To assess the reliability of heterogeneous literature redshifts,
we compared the multiple measurements that are available for some
sources (Fig.\,\ref{fig:redshifts_comp}). \citet{2023A&A...674A...1G}
reported some relatively high redshift measurements that found no
confirmation among other sources (pale blue dots on the left panel).
Many outliers in Fig.\,\ref{fig:redshifts_comp} (left) have a
probability $P_{QSO}$ of being a quasar $\sim$ 100$\%$ questioning
the reliability of P$_{QSO}$ and the values or of their redshitfs.
Further investigation revealed that these high redshifts are
associated with objects not classified as active galaxies or all
together lacking classification in \citet[][again, refer to the
Class in Table\,\ref{tab:redshifts}]{2023A&A...674A...1G}. Therefore,
the redshifts in Gaia\,DR3\, must be treated with caution, unless
the object is specifically classified as a high-probability active
galaxy.

Other literature sources agreed better between themselves. For
example \citet{2024A&A...687A..16I} finds good agreement between
their own redshifts and the Quaia measurements. However, taking
the literature redshifts at face value seems risky. To ensure we
use the most reliable redshifts, we adopted the following strategy:
if an individual redshift measurement for an object is available,
e.g. from our own spectra or from the literature, we prefer
this value, because these as a rule are human-verified, as
opposed to redshifts from large surveys that rely -- as a rule --
on automated procedures. This is largely driven by our concern
about the objects with $z$$>$3 of \citet{2023A&A...674A...1G}. In
total 18 quasars fall into this group. Note that some of the
spectra come from the literature, but we reidentified the emission
lines and remeasured the redshifts.

If no individual measurements were available, we turned to survey
data, in order of preferences: \citet{2023ApJ...944....1D},
\citet{2024ApJ...964...69S}.
%, and \citet{2023A&A...674A...1G}.
These two surveys contributed 9 and 81 redshifts,
respectively. This brings the total number of quasars with
spectroscopic redshifts within the $\mu_B$=26$^m$/$\Box\arcsec$
isophote to 124, omitting the blazar, as discussed earlier.

\section{M\,31 disk extinction}\label{sec:extinction}

We attempted to measure the extinction within the M\,31 disk from
the color excess of quasars within the $\mu_B$=26$^m$/$\Box\arcsec$
isophote with respect to the colors of quasars at similar redshift,
located away from M\,31, assuming that they show on average
the same colors, as long as they are at the same redshift, and that
the large number of quasars away from the disk will average out any
intrinsic variations.

First, we correct both quasar samples for the foreground Milky Way
extinction. Outside of M\,31, we apply individual estimates from
the map of \citet{2011ApJ...737..103S}. Within the M\,31 body, the
predicted E($B$$-$$V$) increases inwards -- due to the dust content of
the galaxy itself. To alleviate this issue, we adopted for all
objects inside the $\mu_B$=26$^m$/$\Box\arcsec$ isophote a constant
foreground Milky Way extinction.
We considered 326 Quaia quasars within an elliptical annulus
between one and two major-axis D26 isophotal diameters, to ensure
that they are safely clear of M\,31, but still in the general
direction of the galaxy, minimizing any residual Milky Way extinction.
Averaging the individual extinctions, we derived
E($B$$-$$V$)=0.060$\pm$0.010\,mag, similar to E($B$$-$$V$)=0.055\,mag
that \citet{2011ApJ...737..103S} report for the M\,31 center and well
within the range 0.035$-$0.105 mag that the same map enables.

\begin{figure*}[t!]
\centering
\includegraphics[width=19.0cm]{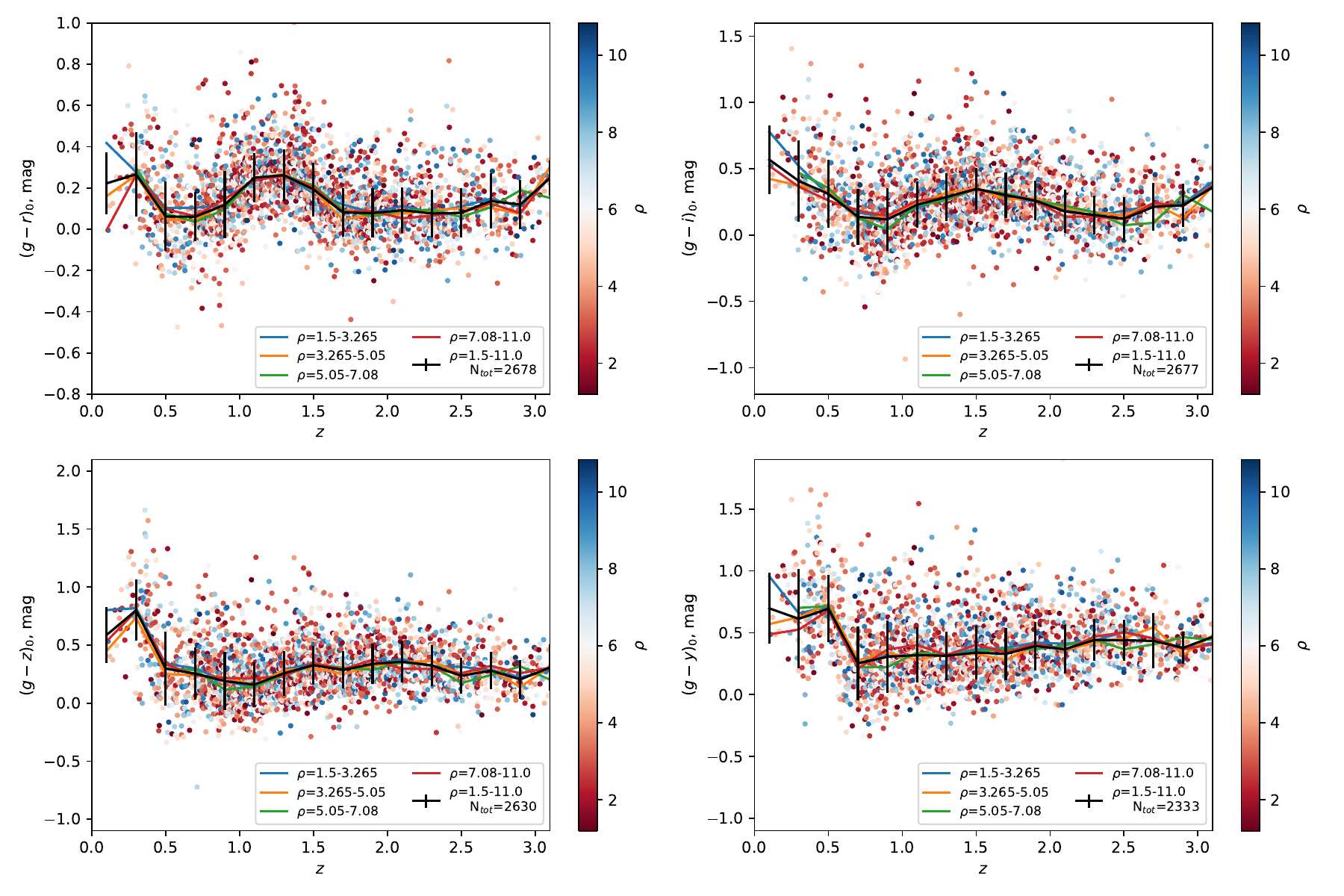}
\caption{Intrinsic colors of our Quaia reference quasar sample,
obtained after correcting the Pan-STARRS Survey DR1 $grizy$ photometry
\citep{2016arXiv161205560C} for the Milky Way extinction according
to the \citet{2011ApJ...737..103S}. The color of the points marks
the distance of the quasars from the M\,31 center $\rho$ in units
of $\rho_{25}$ isophotal radii. The distance bins'
widths were set to ensure these bins contain similar number of
quasars. The solid lines connect the medians colors within 0.5\,dex
wide redshift bins. The vertical error bars are the standard
deviations of the colors of all quasars within each redshift bin,
regardless of the distance from the M\,31 center, and they were
calculated as 1.48$\times$MAD (median absolute deviation), to
suppress the influence of outliers.}
\label{fig:quasar_intrinsic_colors}
\end{figure*}

Next, we derived the intrinsic colors ($g$$-$$r$)$_0$, ($g$$-$$i$)$_0$,
($g$$-$$z$)$_0$ and ($g$$-$$y$)$_0$, at the
redshift of each quasar behind M\,31, as median over the dereddened
colors of all quasars in the reference sample with redshift within
0.3\,dex (Fig.\,\ref{fig:quasar_intrinsic_colors}). This reference
sample was selected from the Quaia catalog to include all quasars
within 5\,deg from the M\,31 center (about 2600 objects in total, see
Fig.\,\ref{fig:quasar_intrinsic_colors} for the exact numbers in the
individual colors), but outside the $\mu_B$=26$^m$/$\Box\arcsec$
isophote, to ensure averaging over 100's of quasars. We
converted the color excesses to A$_V$ adopting a total-to-selective
extinction ratio R$_V$=3.1 and according to the extinction law of
\citet{1989ApJ...345..245C}. We formed the A$_V$ error for each quasar
behind M\,31 as a quadrature sum of the quasars' photometric errors
and the variation of the intrinsic colors for the quasars at that
redshift. We averaged the derived A$_V$'s from the four colors with
error weighting. Some extinctions are negative because of the large
intrinsic spread of quasar colors that can be seen in
Fig.\,\ref{fig:quasar_intrinsic_colors} and to lesser extend, the
photometric errors. Furthermore, undetected contamination from blue
foreground objects in M\,31 itself may also lead to negative
extinction, although we tried to exclude this by selecting isolated
candidates for the follow up. No trend in the evolution of the
broad band intrinsic colors of quasars as a function of redshift is
seen when the sample is divided into several radial bins spanning
1.5 to 11 $\rho_{25}$.\footnote{A stand-alone Python script that
derives the intrinsic colors of quasars as a function of redshift
from a sample of references quasars outside M\,31, and then
calculates the color excesses and absorptions for individual
quasars behind the M\,31, is available at
\url{https://github.com/vdivanov/Quasars_behind_M31_2026/},
together with the necessary input files.}

\begin{figure}[t!]
\centering
\includegraphics[width=9.6cm]{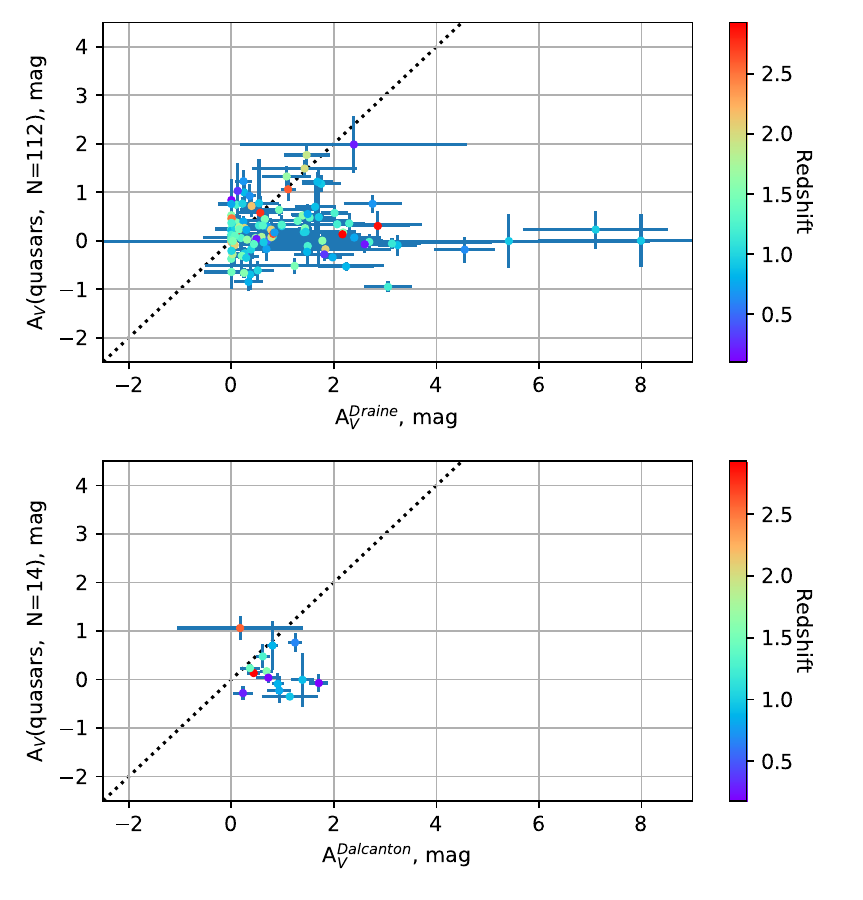}
\caption{Comparison of A$_V$ derived from our quasar sample and from
the reddening maps of \citet[][top]{2014ApJ...780..172D} and
\citet[][bottom]{2015ApJ...814....3D}.}
\label{fig:Av_maps}
\end{figure}

\begin{figure}[t!]
\centering
\includegraphics[width=9.4cm]{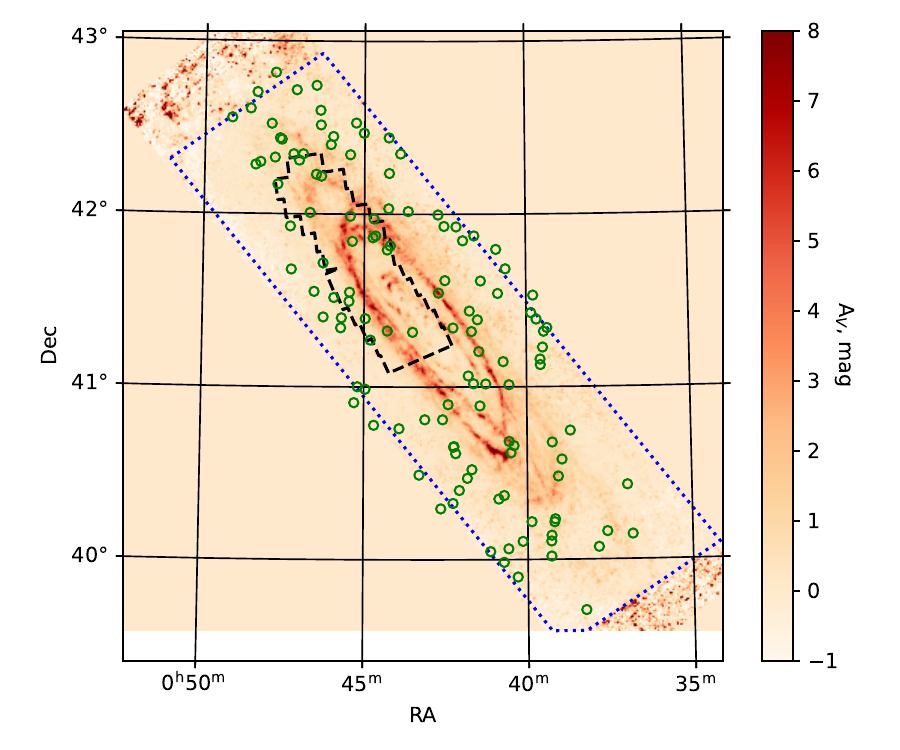}
\caption{Location of the quasars from our sample on the reddening maps
(green open circles) of \citet[][blue dotted line]{2014ApJ...780..172D}
and \citet[][black dashed line]{2015ApJ...814....3D}.}\label{fig:qso_Av_map}
\end{figure}

We used two M\,31 extinction maps to compare our estimates with.
The map of \citet{2014ApJ...780..172D} has the SPIRE 350 resolution
of 25$\arcsec$. It gives the dust surface density and its uncertainty
in units of ($M_{\odot}/kpc^2$). We converted these quantities into
absorption $A_V$ values according to the \citet{2007ApJ...657..810D}
dust model. The map of \citet{2015ApJ...814....3D}  has 7$\arcsec$
resolution. It derives A$_V$ using stellar photometry of red giant
branch (RGB) stars from \citet[PHAT;][]{2014ApJS..215....9W} survey,
enabling a direct comparison with our results. Only 114
of our quasars fall within the footprint of the former map and  14 --
of the latter. No correlation is found in both
cases (Fig.\,\ref{fig:Av_maps}). This negative result is probably due
mostly to the bias towards lower extinction that is rooted in the
quasar selection and especially in the target selection for the
spectroscopic confirmation: quasars located along more heavily obscured
lines of sight are fainter and less likely to have spectra. This can be
seen in Fig.\,\ref{fig:qso_Av_map} that overplots the quasars in the
sample on the \citet{2014ApJ...780..172D} reddening map. Furthermore,
with quasars we measure the extinction along a narrow line of sight,
while the reddening maps average the extinction over wider bins. Last
but not least, the number of quasars may be insufficient for this test,
considering the significant variation of the quasar's intrinsic colors
(Fig.\,\ref{fig:quasar_intrinsic_colors}).

\section{Summary and conclusions}\label{sec:summary}

We selected quasar candidates from apparent colors and X-ray emission
among
objects within the M\,31 $\mu_B$=26$^m$/$\Box\arcsec$ isophote and
carried out optical spectroscopy of the brightest of them to confirm
their nature increasing the number of published spectra from 24 to
40. We also collected quasars from various public surveys, bringing
the total number of bona-fide quasars behind the M\,31 disk to 124
(125, counting one blazar).

Our work provides a large sample of quasars for astrometric studies
of M\,31 \citep[i.e.,][]{2024A&A...692A..30R}, similar to the previous
work in the Magellanic clouds \citep{2011AJ....141..136C}. Quasars
facilitate also studies of its interstellar medium (ISM) -- both of
the extinction from their colors, as we attempted to do here, and of
the ISM's chemical abundances with spectroscopy, typically in the UV
\citep{2013ApJ...772..110F,2024ApJ...976L..28M}. The later work will
come within reach for many faint quasars with the next generation of
30-40-m class spectroscopic telescopes or with the dedicated
spectroscopic facilities, such as the WST \citep[Wide-field
Spectroscopic Telescope;][]{2024arXiv240305398M}.

We measured the extinction inside M\,31 towards 114 quasars
with reliable redshifts and found no correlation with the existing
reddening maps, probably because of the limited number of objects, the
biases in our sample, and different physical scales of the regions
that are probed with the quasars and the dust maps. This problem must
be revisited when LSST helps to discover more quasars, based on their
variability, as \citet{2016A&A...588A..93I,2024A&A...687A..16I}
demonstrated that the variability helps to increase the efficiency of
quasar selection.

Last but not least, in the course of our analysis we confirmed the
previous concerns that the redshifts, derived from low-resolution
spectra may contain significant errors. Such measurements must be
treated with caution.

\begin{acknowledgements}
%\section*{Acknowledgements}
The authors thank an anonymous referee whose valuable comments and
recommendations helped to improve the quality of the paper. PN and AV
acknowledge financial support from the European Union-NextGenerationEU,
through the National Recovery and Resilience Plan of the Republic of
Bulgaria, project No BG-RRP-2.004-0008-C01. PN thanks Phillip Massey
for sharing the spectra of 10 QSOs behind the disk of M\,31 obtained
with Hectospec at the MMT. PN is also grateful to Julianne Dalcanton
for providing the extinction maps of the M\,31 disk, based on the PHAT
survey data. The search for QSOs behind the disc of the M\,31 galaxy
was partially supported by a short-term scholar Fulbright grant of PN
at the Department of Astronomy, University of Washington, for the
academic year 2018–2019 from the US Department of State program number
G-1-00005. This work has made use of data from the European Space
Agency (ESA) mission {\it Gaia} (\url{https://www.cosmos.esa.int/gaia}),
processed by the {\it Gaia} Data Processing and Analysis Consortium
(DPAC, \url{https://www.cosmos.esa.int/web/gaia/dpac/consortium}).
Funding for the DPAC has been provided by national institutions, in
particular the institutions participating in the {\it Gaia} Multilateral
Agreement.
Observations with the SAO RAS telescopes are supported by the Ministry of
Science and Higher Education of the Russian Federation. The renovation of
telescope equipment is currently provided within the national project ``Science
and Universities''. The BTA data reduction was performed as part of the SAO RAS
government contract approved by the Ministry of Science and Higher Education of
the Russian Federation.
This work uses data from the Guoshoujing Telescope (the Large Sky Area
Multi-Object Fiber Spectroscopic Telescope LAMOST). It is a National Major
Scientific Project built by the Chinese Academy of Sciences. Funding for the
project has been provided by the National Development and Reform Commission.
LAMOST is operated and managed by the National Astronomical Observatories,
Chinese Academy of Sciences.
\end{acknowledgements}

\begin{table*}[!ht]
\caption{\small
Sample of 125 QSOs with redshifts based on optical spectrum,
located within M\,31 isophote $\mu_B$=26$^m$/$\Box\arcsec$. The
position and the ID are from Gaia\,DR3 \citep{2023A&A...674A...1G},
except for the objects marked with $^a$; for them this information
comes from Pan-STARRS Survey DR1 \citep{2016arXiv161205560C}. The
distance $\rho$ within the rectified plane of M\,31 is in fractions
of the major isophotal radius $\rho_{25}$. The Class (name of best
class) and the probability of the object being a quasar $P_{QSO}$
(in \%) from the Discrete Source Classifier-Combmode from Gaia DR3
\citep{2023A&A...674A...41A}; when an object is missing a Class in
there, we list its Class as ``uncl'' from unclassified, and if the
object has no entry there, then the Class is set to ``NA'' from
not applicable.
References:
(1) \citet{2023A&A...674A...1G};
(2) \citet{2024ApJ...964...69S};
(3) \citet{2023ApJ...944....1D};
(4) \citet{2013AJ....145..159H};
(5) \citet{2012ApJ...759...11N};
(6) this work;
(7) \citet{2018AJ....155..189D};
(8) \citet{2019RAA....19...29L};
(9) \citet{2019AJ....157..227M};
(10) \citet{1988ApJS...67..249D};
(11) \citet{2010A&A...512A...1M};
(12) \citet{2017ApJ...850...86D};
(13) \citet{2018ATel12250....1N};
(14) \citet{2013MNRAS.432..866R};
(15) \citet{2017ApJ...851..135P}.
Multiple references are given when we have reanalyzed spectra from
the literature to remeasure the redshifts. Finally adopted
redshift $z_{adopt}$ is explained in Sect. 4.3. The analyzed original
spectra are suffixed by A (Apache Point) or B (BTA) in the last
column while the reanalyzed literature spectra are suffixed by M
(Mayall) or L (LAMOST).}\label{tab:redshifts}
\vspace{-2mm}
{\footnotesize
\begin{tabular}{@{}l@{ }c@{ }p{1.5cm}c@{}c@{ }c@{ }c@{ }c@{ }p{0.5cm}c@{ }p{1.7cm}p{0.8cm}p{0.7cm}c@{}c@{}}
\hline
No& Gaia\,DR3        & ~~~RA DEC          &~~~$\rho$~~~&~Gmag,~ &~~rms,~~&~Class~&P\%  &~$z_{lit}$   &~Ref.~~& ~Emis. & $z_{adopt}$ & $A_V$ &Obs.     & ~~Fig.\,3   \\
  & Name             & ~~~~~J2000         &            &  mag   & mag    &       &~QSO~&             &       & ~lines    &             & mag   &log\,ID  & ~~ID \\
\hline
  1 &~380728681624635264~   & 00:36:49.070 +40:08:41.73 & 1.161 & 18.921 & 0.078 & uncl & 100 & 0.502$\pm$ 0.051   &  2  &  & 0.502$\pm$ 0.051   &    0.000$\pm$ 0.134 &  &  \\
  2 & 380803551495086592    & 00:36:57.656 +40:25:50.28 & 1.228 & 19.311 & 0.095 &  AGN &  97 & 1.181$\pm$ 0.019   &  1  &  & 1.208$\pm$ 0.005   & $-$0.262$\pm$ 0.076 &  &  \\
    &                       &                           &       &        &       &      &     & 1.181$\pm$ 0.074   &  2  &  &                    &                     &  &  \\
    &                       &                           &       &        &       &      &     & 1.208              &  3  &  &                    &                     &  &  \\
  3 & 380749950302728832    & 00:37:35.058 +40:09:41.68 & 0.997 & 18.682 & 0.114 &  AGN & 100 & 0.337$\pm$ 0.001   &  1  &  & 0.337$\pm$ 0.085   &    0.040$\pm$ 0.129 &  &  \\
    &                       &                           &       &        &       &      &     & 0.337$\pm$ 0.085   &  2  &  &                    &                     &  &  \\
\hline
\end {tabular}
}
\end {table*}

\addtocounter{table}{-1}
\begin{table*}[!ht]
\caption{Continued.}
\vspace{-3mm}
{\footnotesize
\begin{tabular}{@{}l@{ }c@{ }p{1.5cm}c@{}c@{ }c@{ }c@{ }c@{ }p{0.5cm}c@{ }p{1.7cm}p{0.8cm}p{0.7cm}c@{}c@{}}
\hline
No& Gaia\,DR3        & ~~~RA DEC          &~~~$\rho$~~~&~Gmag,~ &~~rms,~~&~Class~&P\%  &~$z_{lit}$   &~Ref.~~& ~Emis. & $z_{adopt}$ & $A_V$ &Obs.     & ~~Fig.\,3   \\
  & Name             & ~~~~~J2000         &            &  mag   & mag    &       &~QSO~&             &       & ~lines    &             & mag   &log\,ID  & ~~ID \\
\hline
 14 & 368728508640097792    & 00:39:17.049 +40:01:06.97 & 0.957 & 20.328 & 0.173 &  AGN & 100 & 2.046$\pm$ 0.016   &  1  &  & 2.046$\pm$ 0.074   &    0.228$\pm$ 0.226 &  &  \\
    &                       &                           &       &        &       &      &     & 2.046$\pm$ 0.074   &  2  &  &                    &                     &  &  \\
 15 & 381231021002898048    & 00:39:21.921 +41:20:31.31 & 1.219 & 20.525 & 0.277 &  AGN & 100 & 4.619$\pm$ 0.042   &  1  &  & 1.425$\pm$ 0.005   & $-$0.644$\pm$ 0.347 &  &  \\
    &                       &                           &       &        &       &      &     & 1.425$\pm$ 0.005   &  6  & C{\sc iii}]\,1909, Mg{\sc ii}\,2799 & & &1 & 3921+4120B \\ %BTA Yao QSOcand1 \\
 16 & 381230200663395712    & 00:39:28.105 +41:19:13.09 & 1.160 & 20.765 & 0.224 &  AGN &  69 & 2.029$\pm$ 0.025   &  1  &  & 2.057$\pm$ 0.001   & $-$0.176$\pm$ 0.181 &  &  \\
    &                       &                           &       &        &       &      &     & 2.057$\pm$ 0.001   &  6  & C{\sc iv}\,1548, C{\sc iii}]\,1909 & & &2  & 3928+4119B \\
\hline
\end {tabular}
}
\end {table*}

\addtocounter{table}{-1}
\begin{table*}[!ht]
\caption{Continued.}
\vspace{-3mm}
{\footnotesize
\begin{tabular}{@{}l@{ }c@{ }p{1.5cm}c@{}c@{ }c@{ }c@{ }c@{ }p{0.5cm}c@{ }p{1.7cm}p{0.8cm}p{0.7cm}c@{}c@{}}
\hline
No& Gaia\,DR3        & ~~~RA DEC          &~~~$\rho$~~~&~Gmag,~ &~~rms,~~&~Class~&P\%  &~$z_{lit}$   &~Ref.~~& ~Emis. & $z_{adopt}$ & $A_V$ &Obs.     & ~~Fig.\,3   \\
  & Name             & ~~~~~J2000         &            &  mag   & mag    &       &~QSO~&             &       & ~lines    &             & mag   &log\,ID  & ~~ID \\
\hline
 27 & 381129011231416704    & 00:40:29.727 +40:37:05.68 & 0.491 & 18.986 & 0.076 &   NA &   - & 1.585$\pm$ 0.007   &  6  & Si{\sc v}\,1400\,C{\sc iii}] 1909,\,Mg{\sc ii}\,2799 & 1.585$\pm$ 0.007   &    0.521$\pm$ 0.077 & 6 & 4029+4037A \\ %APO W66\\
 28 & 381172953042364032    & 00:40:32.571 +41:00:43.81 & 0.500 & 20.001 & 0.144 &  AGN &  15 & 0.358$\pm$ 0.011   &  1  &  & 0.845$\pm$ 0.220   &    1.216$\pm$ 0.201 &  &  \\
    &                       &                           &       &        &       &      &     & 0.845$\pm$ 0.220   &  2  &  &                    &                     &  &  \\
 29 & 381130007666198272    & 00:40:32.954 +40:41:02.66 & 0.451 & 19.451 & 0.092 & uncl &  38 & 4.347$\pm$ 0.009   &  1  &  & 0.944$\pm$ 0.180   &    0.230$\pm$ 0.390 &  &  \\
    &                       &                           &       &        &       &      &     & 0.944$\pm$ 0.180   &  2  &  &                    &                     &  &  \\
 30 & 369089496347724160    & 00:40:35.038 +40:03:45.04 & 1.080 & 19.951 & 0.125 &  AGN &  74 & 1.345$\pm$ 0.016   &  1  &  & 1.463$\pm$ 0.368   &    0.105$\pm$ 0.133 &  &  \\
\hline
\end {tabular}
}
\end{table*}

\addtocounter{table}{-1}
\begin{table*}[!ht]
\caption{Continued.}
\vspace{-3mm}
{\footnotesize
\begin{tabular}{@{}l@{ }c@{ }p{1.5cm}c@{}c@{ }c@{ }c@{ }c@{ }p{0.5cm}c@{ }p{1.7cm}p{0.8cm}p{0.7cm}c@{}c@{}}
\hline
No& Gaia\,DR3        & ~~~RA DEC          &~~~$\rho$~~~&~Gmag,~ &~~rms,~~&~Class~&P\%  &~$z_{lit}$   &~Ref.~~& ~Emis. & $z_{adopt}$ & $A_V$ &Obs.     & ~~Fig.\,3   \\
  & Name             & ~~~~~J2000         &            &  mag   & mag    &       &~QSO~&             &       & ~lines    &             & mag   &log\,ID  & ~~ID \\
\hline
    &                       &                           &       &        &       &      &     & 0.628$\pm$ 0.059   &  2  &  &                    &                     &  &  \\
 43 & 381277544086058880    & 00:41:30.240 +41:23:20.22 & 0.555 & 19.505 & 0.085 &   NA &  -  & 1.895              &  9  &  & 1.900$\pm$ 0.011   &    1.768$\pm$ 0.246 &  &  \\  %MMT J004130.25+412320.3\\\
    &                       &                           &       &        &       &      &     & 1.900$\pm$ 0.011   & 6;9 & C{\sc iii}]\,1909 & & & & 4130+4123M \\ %MMT J004130.25+412320.3 other features affected by intervening absorption\\
 44 & 381341251346342016    & 00:41:36.120 +41:52:29.22 & 1.141 & 20.291 & 0.226 & uncl &  69 & 1.192$\pm$ 0.024   &  1  &  & 1.192$\pm$ 0.607   &    0.537$\pm$ 0.119 &  &  \\
    &                       &                           &       &        &       &      &     & 1.192$\pm$ 0.607   &  2  &  &                    &                     &  &  \\
 45 & 381164710999391104    & 00:41:37.918 +41:01:07.82 & 0.213 & 19.404 & 0.091 & uncl &  79 & 1.450              &  5  &  & 1.466$\pm$ 0.002   &    0.445$\pm$ 0.154 &  &  \\
\hline
\end {tabular}
}
\end{table*}

\addtocounter{table}{-1}
\begin{table*}[!ht]
\caption{Continued.}
\vspace{-3mm}
{\footnotesize
\begin{tabular}{@{}l@{ }c@{ }p{1.5cm}c@{}c@{ }c@{ }c@{ }c@{ }p{0.5cm}c@{ }p{1.7cm}p{0.8cm}p{0.7cm}c@{}c@{}}
\hline
No& Gaia\,DR3        & ~~~RA DEC          &~~~$\rho$~~~&~Gmag,~ &~~rms,~~&~Class~&P\%  &~$z_{lit}$   &~Ref.~~& ~Emis. & $z_{adopt}$ & $A_V$ &Obs.     & ~~Fig.\,3   \\
  & Name             & ~~~~~J2000         &            &  mag   & mag    &       &~QSO~&             &       & ~lines    &             & mag   &log\,ID  & ~~ID \\
\hline
    &                       &                           &       &        &       &      &     & 0.7579$\pm$ 0.0007~ & ~~6;8 & Mg{\sc ii}\,2799,\,O{\sc ii} 3727,\.H$\beta$\,4863, O{\sc iii}\,4959\&5007 & & & & 4214.1+4039L \\ %spec-57013-M31011N40M2\_sp04-174.fits LAM DR5\\
 56 & 369145773303561856    & 00:42:14.991 +40:39:09.54 & 0.690 & 19.874 & 0.113 &  AGN &  46 & 1.780$\pm$ 0.010   &  1  &  & 1.778$\pm$ 0.001   &    0.079$\pm$ 0.034 &  &  \\
    &                       &                           &       &        &       &      &     & 1.774              &  9  &  &                    &                     &  &  \\   %MMT J004214.98+403909.8\\
    &                       &                           &       &        &       &      &     & 1.778$\pm$ 0.001   & 6;9 & C{\sc iv}\,1548, C{\sc iii}]\,1909 & & & & 4214+4039M \\ %MMT J004214.98+403909.8\\
 57 & 381266244030810880    & 00:42:15.487 +41:20:31.52 & 0.249 & 20.549 & 0.156 &   NA &   - & 1.520$\pm$ 0.020   &  6  & C{\sc iv}\,1548,\,C{\sc iii}] 1909,\,Mg{\sc ii}\,2799 & 1.520$\pm$ 0.020   &    0.259$\pm$ 0.400 & 11 & 4215+4120A \\ %APO W237\\
 58 & 369120145234465024    & 00:42:16.121 +40:19:35.88 & 1.127 & 16.852 & 0.326 &  AGN & 100 & 0.101$\pm$ 0.001   &  1  &  & 0.1029$\pm$ 0.0002 &    0.838$\pm$ 0.425 &  &  \\
\hline
\end {tabular}
}
\end{table*}

\addtocounter{table}{-1}
\begin{table*}[!ht]
\caption{Continued.}
\vspace{-3mm}
{\footnotesize
\begin{tabular}{@{}l@{ }c@{ }p{1.5cm}c@{}c@{ }c@{ }c@{ }c@{ }p{0.5cm}c@{ }p{1.7cm}p{0.8cm}p{0.7cm}c@{}c@{}}
\hline
No& Gaia\,DR3        & ~~~RA DEC          &~~~$\rho$~~~&~Gmag,~ &~~rms,~~&~Class~&P\%  &~$z_{lit}$   &~Ref.~~& ~Emis. & $z_{adopt}$ & $A_V$ &Obs.     & ~~Fig.\,3   \\
  & Name             & ~~~~~J2000         &            &  mag   & mag    &       &~QSO~&             &       & ~lines    &             & mag   &log\,ID  & ~~ID \\
\hline
 70 & 387338258335949952    & 00:43:52.900 +42:20:58.01 & 1.149 & 18.902 & 0.131 &  AGN &  80 & 0.774$\pm$ 0.009   &  1  &  & 0.764$\pm$ 0.152   &    0.758$\pm$ 0.246 &  &  \\
    &                       &                           &       &        &       &      &     & 0.764$\pm$ 0.152   &  2  &  &                    &                     &  &  \\
 71 & 369227102806068736    & 00:43:55.044 +40:45:32.81 & 1.033 & 19.768 & 0.146 &  AGN &  99 & 0.860$\pm$ 0.009   &  1  &  & 0.866$\pm$ 0.005   & $-$0.697$\pm$ 0.317 &  &  \\
    &                       &                           &       &        &       &      &     & 0.860$\pm$ 0.064   &  2  &  &                    &                     &  &  \\
    &                       &                           &       &        &       &      &     & 0.866              &  3  &  &                    &                     &  &  \\
 72 & 381307024742647168    & 00:44:11.968 +41:49:01.33 & 0.429 & 19.298 & 0.082 & uncl &  12 & 1.087$\pm$ 0.013   &  1  &  & 0.747$\pm$ 0.120   &   0.065$\pm$ 0.149  &  &  \\
\hline
\end {tabular}
}
\end{table*}

\addtocounter{table}{-1}
\begin{table*}[!ht]
\caption{Continued.}
\vspace{-3mm}
{\footnotesize
\begin{tabular}{@{}l@{ }c@{ }p{1.5cm}c@{}c@{ }c@{ }c@{ }c@{ }p{0.5cm}c@{ }p{1.7cm}p{0.8cm}p{0.7cm}c@{}c@{}}
\hline
No& Gaia\,DR3        & ~~~RA DEC          &~~~$\rho$~~~&~Gmag,~ &~~rms,~~&~Class~&P\%  &~$z_{lit}$   &~Ref.~~& ~Emis. & $z_{adopt}$ & $A_V$ &Obs.     & ~~Fig.\,3   \\
  & Name             & ~~~~~J2000         &            &  mag   & mag    &       &~QSO~&             &       & ~lines    &             & mag   &log\,ID  & ~~ID \\
\hline
 85 & 387349012934217856    & 00:45:01.357 +42:28:07.73 & 1.068 & 20.333 & 0.144 &  AGN &   2 & 1.680$\pm$ 0.017   &  1  &  & 2.588$\pm$ 0.743   &    1.061$\pm$ 0.249 &  &  \\
    &                       &                           &       &        &       &      &     & 2.588$\pm$ 0.743   &  2  &  &                    &                     &  &  \\
 86 & 369212809151892224    & 00:45:11.917 +41:00:04.05 & 1.141 & 19.109 & 0.095 &  AGN &  98 & 1.125$\pm$ 0.016   &  1  &  & 1.125$\pm$ 0.080   & $-$0.059$\pm$ 0.115 &  &  \\
    &                       &                           &       &        &       &      &     & 1.125$\pm$ 0.080   &  2  &  &                    &                     &  &  \\
    &                       &                           &       &        &       &      &     & 1.144              &  4  &  &                    &                     &  &  \\
 87 & 387355232046935808    & 00:45:16.119 +42:31:38.89 & 1.098 & 20.063 & 0.121 &  AGN &  53 & 1.587$\pm$ 0.019   &  1  &  & 1.596$\pm$ 0.005   &    0.053$\pm$ 0.034 &  &  \\
\hline
\end {tabular}
}
\end{table*}

\addtocounter{table}{-1}
\begin{table*}[!ht]
\caption{Continued.}
\vspace{-3mm}
{\footnotesize
\begin{tabular}{@{}l@{ }c@{ }p{1.5cm}c@{}c@{ }c@{ }c@{ }c@{ }p{0.5cm}c@{ }p{1.7cm}p{0.8cm}p{0.7cm}c@{}c@{}}
\hline
No& Gaia\,DR3        & ~~~RA DEC          &~~~$\rho$~~~&~Gmag,~ &~~rms,~~&~Class~&P\%  &~$z_{lit}$   &~Ref.~~& ~Emis. & $z_{adopt}$ & $A_V$ &Obs.     & ~~Fig.\,3   \\
  & Name             & ~~~~~J2000         &            &  mag   & mag    &       &~QSO~&             &       & ~lines    &             & mag   &log\,ID  & ~~ID \\
\hline
 99 & 369271740398774784    & 00:46:15.916 +41:24:14.62 & 1.037 & 19.789 & 0.142 &  AGN & 100 & 0.946$\pm$ 0.011   &  1  &  & 0.946$\pm$ 0.053   & $-$0.352$\pm$ 0.044 &  &  \\
    &                       &                           &       &        &       &      &     & 0.946$\pm$ 0.053   &  2  &  &                    &                     &  &  \\
100 & 375284415439911296    & 00:46:16.832 +41:43:00.73 & 0.767 & 20.479 & 0.156 & uncl &  69 & 5.883$\pm$ 0.057   &  1  &  & 0.974$\pm$ 0.246   &    0.272$\pm$ 0.211 &  &  \\
    &                       &                           &       &        &       &      &     & 0.974$\pm$ 0.246   &  2  &  &                    &                     &  &  \\
101 & 375332759593218432    & 00:46:21.066 +42:13:08.00 & 0.734 & 17.846 & 0.076 & uncl & 100 & 4.350$\pm$ 0.004   &  1  &  & 0.920$\pm$ 0.053   & $-$0.032$\pm$ 0.213 &  &  \\
    &                       &                           &       &        &       &      &     & 0.920$\pm$ 0.053   &  2  &  &                    &                     &  &  \\
\hline
\end {tabular}
}
\end{table*}

\addtocounter{table}{-1}
\begin{table*}[!ht]
\caption{Continued.}
\vspace{-3mm}
{\footnotesize
\begin{tabular}{@{}l@{ }c@{ }p{1.5cm}c@{}c@{ }c@{ }c@{ }c@{ }p{0.5cm}c@{ }p{1.7cm}p{0.8cm}p{0.7cm}c@{}c@{}}
\hline
No& Gaia\,DR3        & ~~~RA DEC          &~~~$\rho$~~~&~Gmag,~ &~~rms,~~&~Class~&P\%  &~$z_{lit}$   &~Ref.~~& ~Emis. & $z_{adopt}$ & $A_V$ &Obs.     & ~~Fig.\,3   \\
  & Name             & ~~~~~J2000         &            &  mag   & mag    &       &~QSO~&             &       & ~lines    &             & mag   &log\,ID  & ~~ID \\
\hline
113 & 375313213198402432    & 00:47:18.499 +41:55:44.21 & 0.936 & 19.723 & 0.113 &  AGN &   8 & 0.752$\pm$ 0.021   &  1  &  & 1.806$\pm$ 0.509   &    0.278$\pm$ 0.246 &  &  \\
    &                       &                           &       &        &       &      &     & 1.806$\pm$ 0.509   &  2  &  &                    &                     &  &  \\
114 & 375423920274397824    & 00:47:35.750 +42:25:49.99 & 0.937 & 20.402 & 0.189 & uncl &  71 & 5.052$\pm$ 0.057   &  1  &  & 1.564$\pm$ 0.274   &    0.001$\pm$ 0.209 &  &  \\
    &                       &                           &       &        &       &      &     & 1.564$\pm$ 0.274   &  2  &  &                    &                     &  &  \\
115 & 375423881617811328    & 00:47:39.120 +42:26:17.76 & 0.945 & 18.197 & 0.065 &  AGN & 100 & 2.880$\pm$ 0.026   &  1  &  & 0.705$\pm$ 0.001   &    0.171$\pm$ 0.136 &  &  \\
    &                       &                           &       &        &       &      &     & 0.850$\pm$ 0.147   &  2  &  &                    &                     &  &  \\
\hline
\end {tabular}
}
\end{table*}

\begin{appendix}
%\onecolumn
\section{Data availability}\label{app:data_availab}

Table\,\ref{tab:data_spectra} lists the spectra that are reported for
the first time in this paper.

\begin{table}[!ht]
\begin{center}
\caption{New spectra, reported here for the first time. Only part
of the table is shown as an example. The entire table is available
only in electronic form at the CDS.}\label{tab:data_spectra}
\vspace{-2mm}
\begin{tabular}{c@{ }c}
\hline
$\lambda$ [Angstr\"om]~ & ~~Flux [arbitrary units]\\
\hline
\multicolumn{2}{c}{J004029.727+403705.68}\\
        3533   &    1.104 \\
        3535   &    1.222 \\
        3537   &    0.995 \\
        3538   &    1.229 \\
        3540   &    1.041 \\
        3542   &    0.960 \\
        3544   &    1.293 \\
\multicolumn{2}{c}{...}\\
\multicolumn{2}{c}{J004215.487+412031.52}\\
        3801  &     0.429 \\
        3803  &     1.196 \\
        3805  &     1.275 \\
        3807  &     1.742 \\
        3809  &     0.767 \\
        3811  &     1.595 \\
        3813  &     2.027 \\
\multicolumn{2}{c}{...}\\
\hline
\end{tabular}
\end{center}
\end{table}
\vspace{-8mm}

\section{Intrinsic colors of quasars}\label{app:data_quasar_colors}

{Table\,\ref{tab:quasar_colors} lists the derived intrinsic colors
of quasars,
as shown with black lines in Fig.\,\ref{fig:quasar_intrinsic_colors}.

\begin{table}[!ht]
\begin{center}
\caption{Intrinsic colors of quasars. N is the number of quasars in
the respective redshift $z$ bin (bin size 0.2\,dex); the numbers of
some colors can be smaller if there are quasars with missing
photometry.}\label{tab:quasar_colors}
\vspace{-2mm}
\begin{tabular}{@{}c@{ }c@{ }c@{ }c@{ }c@{ }r@{}}
\hline
$z$ & ~($g$$-$$r$) [mag]~ & ~($g$$-$$i$) [mag]~ & ~($g$$-$$z$) [mag]~ & ~($g$$-$$y$) [mag]~ & ~~N \\
\hline
0.1 & 0.222$\pm$0.151 & 0.569$\pm$0.258 & 0.586$\pm$0.239 & 0.697$\pm$0.286 &    8 \\
0.3 & 0.266$\pm$0.201 & 0.408$\pm$0.305 & 0.800$\pm$0.264 & 0.612$\pm$0.402 &   79 \\
0.5 & 0.062$\pm$0.168 & 0.312$\pm$0.258 & 0.297$\pm$0.321 & 0.694$\pm$0.272 &  138 \\
0.7 & 0.061$\pm$0.120 & 0.138$\pm$0.213 & 0.254$\pm$0.224 & 0.251$\pm$0.298 &  187 \\
0.9 & 0.118$\pm$0.163 & 0.118$\pm$0.235 & 0.190$\pm$0.251 & 0.306$\pm$0.294 &  249 \\
1.1 & 0.250$\pm$0.118 & 0.229$\pm$0.173 & 0.159$\pm$0.193 & 0.319$\pm$0.223 &  273 \\
1.3 & 0.262$\pm$0.124 & 0.287$\pm$0.182 & 0.257$\pm$0.193 & 0.314$\pm$0.197 &  270 \\
1.5 & 0.193$\pm$0.130 & 0.349$\pm$0.160 & 0.327$\pm$0.167 & 0.339$\pm$0.216 &  295 \\
1.7 & 0.081$\pm$0.116 & 0.301$\pm$0.174 & 0.286$\pm$0.160 & 0.329$\pm$0.210 &  262 \\
1.9 & 0.078$\pm$0.119 & 0.260$\pm$0.148 & 0.337$\pm$0.174 & 0.392$\pm$0.196 &  222 \\
2.1 & 0.092$\pm$0.111 & 0.178$\pm$0.162 & 0.357$\pm$0.179 & 0.367$\pm$0.197 &  173 \\
2.3 & 0.076$\pm$0.116 & 0.151$\pm$0.145 & 0.325$\pm$0.177 & 0.442$\pm$0.171 &  174 \\
2.5 & 0.080$\pm$0.121 & 0.123$\pm$0.162 & 0.235$\pm$0.157 & 0.440$\pm$0.164 &  124 \\
2.7 & 0.136$\pm$0.134 & 0.213$\pm$0.178 & 0.280$\pm$0.161 & 0.434$\pm$0.223 &   94 \\
2.9 & 0.119$\pm$0.119 & 0.225$\pm$0.163 & 0.204$\pm$0.165 & 0.377$\pm$0.131 &   62 \\
3.1 & 0.248$\pm$0.085 & 0.361$\pm$0.223 & 0.306$\pm$0.201 & 0.467$\pm$0.160 &   34 \\
\hline
\end{tabular}
\end{center}
\end{table}
}

\vspace{-8mm}
\section{Correction of the {\it Spitzer} coordinates}\label{app:mid_ir_coords}

\begin{figure}
\includegraphics[width=9cm]{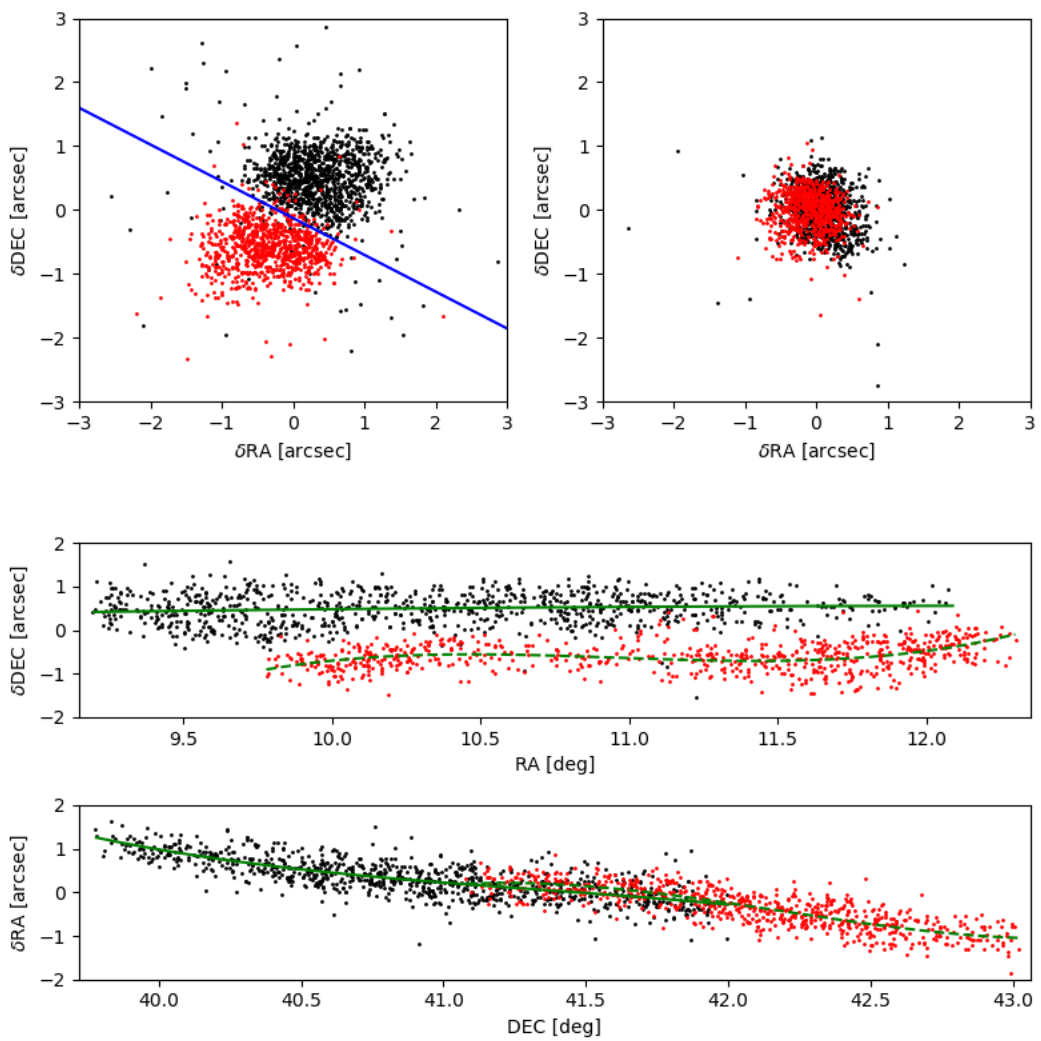}
\caption{Upper left: $\delta$RA/$\delta$DEC plot for the astrometric
{\it Spitzer}--{\it WISE} comparison of 1623 objects from our bright
astrometric sample. The double central clump is due to small
systematic deviations in the \citet{2017ApJS..228....5K} coordinate
system. The blue line makes initial division of the astrometric
sample into two spatially distinct sub-samples of $\sim$700 and
$\sim$900 objects (see Fig.\,\ref{fig:astro_radec}). Upper right:
the same plot after 3$\sigma$ clipping and correcting for the
systematic effect to the WISE coordinate system. Middle:
$\delta$DEC({\it Spitzer}--{\it WISE}) versus {\it WISE} RA. Bottom:
$\delta$RA({\it Spitzer}--{\it WISE}) versus {\it WISE} DEC. The
solid and dashed green curves are the best fits to the trends for
the two sub-samples on the middle and bottom panels (see text).}
\label{fig:quasar_delta-delta}
\end{figure}

\begin{figure}
\centering
\includegraphics[width=8.7cm]{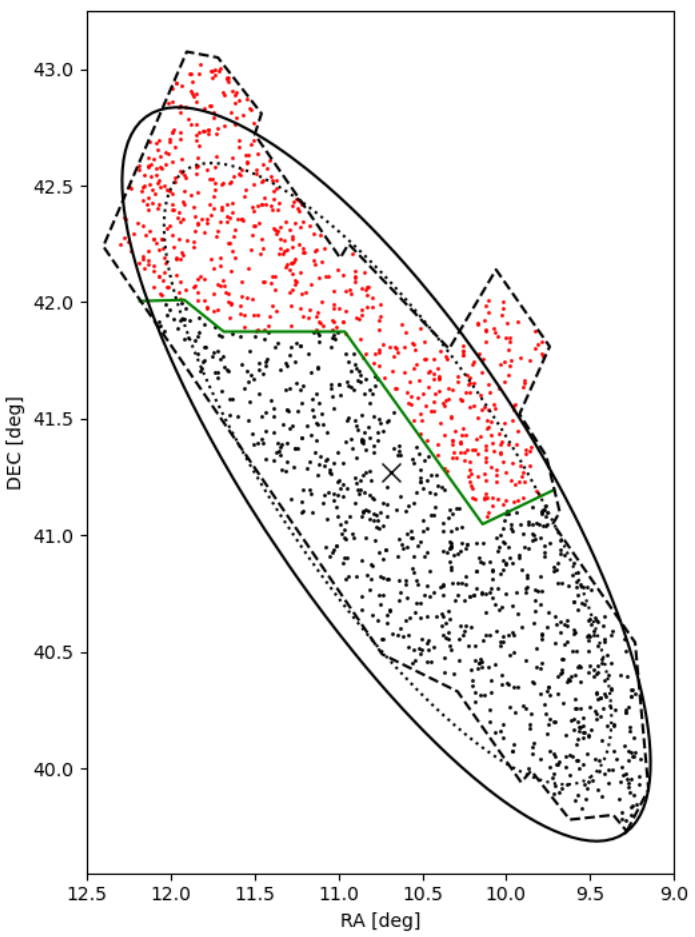}
\caption{Map of our astrometric sample of 1623 objects with the
adopted final dividing solid line (solid green) for the two
spatially distinct samples of 710 (red dots) and 913 (black dots)
objects. The dashed line shows the approximate outline of
\citet{2017ApJS..228....5K} catalog, the solid and dotted lines
show the $\mu_B$=26$^m$/$\Box\arcsec$ and
$\mu_B$=25$^m$/$\Box\arcsec$ isophotes, respectively. M\,31 center
is marked with an x.}
\label{fig:astro_radec}
\end{figure}

The coordinate accuracy of candidate quasars is crucial for their
cross-identification for the optical spectroscopic follow up. Most of
our candidates are selected on the base of their {\it Spitzer}
photometry \citep{2017ApJS..228....5K}, but a comparison with
various near-infrared and optical catalogs revealed systematic offsets
in this coordinate system. Therefore, we decided to transfer the {\it
Spitzer} coordinates of \citet{2017ApJS..228....5K} to the {\it WISE}
J2000 coordinate system. We cross-matched the {\it Spitzer} and
{\it WISE} coordinates with 3$\arcsec$ matching radius, for
objects with [W2]$\leq$13.5\,mag and
--0.35\,mag$\leq$[W2]--[W1]$\leq$0.35\,mag as shown in
Fig.\,\ref{fig:quasar_ccd_cmd_W} (top right panel), to ensure we only
consider high signal-o-noise detections and to reduce the number of
spurious matches. Then, we formed the {\it Spitzer}--{\it WISE}
coordinate differences $\delta$RA and $\delta$DEC, for these sources,
plotted them versus each other (Fig.\ref{fig:quasar_delta-delta}, top
left) and saw a two-clump structure. We imposed an initial separation
between the two clumps along the blue line. The objects above and below it
(plotted everywhere with black and red dots, respectively)
are located into two distinct spatial regions on the sky, separated
with a line connecting the following positions in RA, DEC (J2000), in
degrees: (9.714, 41.195), (10.140, 41.048), (10.963, 41.874), (11.685, 41.874),
(11.920, 42.010) and (12.186, 42.005) -- marked with green solid line in
Fig.\,\ref{fig:astro_radec}. For all objects within each of these two
areas we derived separately polynomial fits, with
3-$\sigma$ clipping, of $\delta$RA and $\delta$DEC versus DEC and RA,
respectively (Fig.\ref{fig:quasar_delta-delta}, middle and lower
panels), based on about 1600 objects in total:
\begin{equation}
\delta DEC_{black}=-1.671+0.36211\times RA-0.014715\times RA^2
\end{equation}
\begin{equation}
\begin{split}
\delta DEC_{red}=-405.401+107.03023\times RA-9.006573\times RA^2\\
+0.2022619\times RA^3+0.00327904\times RA^4
\end{split}
\end{equation}
\begin{equation}
\begin{split}
\delta RA_{black}=-26974.840+1920.434447\times DEC \\
-45.5512001\times DEC^2+0.35995935\times DEC^3
\end{split}
\end{equation}
\begin{equation}
\begin{split}
\delta RA_{red}=8571.288-620.935448\times DEC \\
+15.0056471\times DEC^2 -0.12096752\times DEC^3
\end{split}
\end{equation}
The typical RMS of these relations is $\sim$0.3\arcsec. Applying them
brings the two clumps together (Fig.\ref{fig:quasar_delta-delta},
upper right panel).

A minor caveat in this correction is the omission of the proper
motions of the brightest sources that are probably Milky Way
stars, but the
Gaia\,DR3 suggests there are of order of 10-20\,mas\,yr$^{-1}$.
Fortuitously, the {\it Spitzer}--{\it WISE} time baseline is
$\sim$ 5 yr , so any cumulative
effect from the proper motions can safely be neglected.

\end{appendix}
\end{document}